\documentclass[draft]{agujournal2019}

\usepackage{sah_shortcuts}
\usepackage{amsmath,amssymb}

\usepackage{url} %this package should fix any errors with URLs in refs.
\usepackage{lineno}
\usepackage[inline]{trackchanges} %for better track changes. finalnew option will compile document with changes incorporated.
\usepackage{soul}
\draftfalse

\journalname{Geophysical Research Letters}

\begin{document}

% \title{On the all-India rainfall index, sub-India rainfall heterogeneity, and Indo-Pacific teleconnections}
\title{On the all-India rainfall index and sub-India rainfall heterogeneity}

\authors{Spencer A. Hill\affil{1}, Adam H. Sobel\affil{1,2}, Michela Biasutti\affil{1}, and Mark A. Cane\affil{1}}
\affiliation{1}{Lamont-Doherty Earth Observatory, Columbia University, Palisades, New York}
\affiliation{2}{Department of Applied Physics and Applied Mathematics and Lamont-Doherty Earth Observatory, Columbia University, New York, New York}
\correspondingauthor{Spencer A. Hill}{shill@ldeo.columbia.edu}

%% Keypoints, final entry on title page.
%  List up to three key points (at least one is required)
%  Key Points summarize the main points and conclusions of the article
%  Each must be 100 characters or less with no special characters or punctuation and must be complete sentences
% \item This line ends at exactly 100 characters after the "\item " so don't go past it to stay in bounds!!!
\begin{keypoints}
\item We examine Indian summer monsoon rainfall interannual variability and teleconnections 1901-2020
\item All-India Rainfall Index (AIRI) closely tracks spatial and temporal extent of wet anomalies
\item ENSO teleconnection projects onto AIRI but Indian Ocean Dipole onto a known tripole pattern
\end{keypoints}

% 150 word limit for GRL; 250 for others
\begin{abstract}
We revisit long-standing controversies regarding relationships among the all-India rainfall index (AIRI), sub-India summer rainfall variations, El Ni\~no-Southern Oscillation (ENSO), and the Indian Ocean Dipole (IOD) using 120-year sea surface temperature and high-resolution rainfall datasets.  AIRI closely tracks with the spatial extent of wet anomalies and with the average across gridpoints in rainy day count.  The leading rainfall variability mode is a monopole associated primarily with rainy day count and ENSO.  The second mode is a tripole with same-signed loadings in the high-rainfall Western Ghats and Central Monsoon Zone regions and opposite-signed loadings in Southeastern India between.  The IOD projects onto this tripole and, as such, is weakly correlated with AIRI.  However, when the linear influence of ENSO is removed, the IOD rainfall regressions become quasi-homogeneously more positive, making the ENSO-residual IOD and AIRI timeseries significantly correlated.
% ; it is less strongly related to AIRI and has comparable contributions from rainy-day frequency and intensity.
%by decomposing June-July-August-September (JJAS) rainfall anomalies into the product of the frequency of rainy days and mean rainfall intensity on rainy days. %and assess teleconnections from the tropical Pacific Ocean (via El Ni\~no-Southern Oscillation, ENSO) and Indian Ocean (via the Indian Ocean Dipole, IOD).
% During the Indian summer monsoon season of June-July-August-September (JJAS), developing \elnino/ events in the Pacific Ocean tend to reduce all-India rainfall (AIR), but the strength of this teleconnection varies on decadal timescales.  The influence of Indian Ocean sea surface temperatures (SSTs) on the monsoon rainfall in general and these variations in the \enso/ (ENSO) teleconnection is contested, with both the mean Indian Ocean SST and the zonal gradient along the equator (the Indian Ocean Dipole) being posited as important.
% A widely used indicator of Indian summer monsoon variability is the June-July-August-September (JJAS) rainfall rate averaged over all of India (AIR).  But AIR's relationship with the spatial extent of dry or wet anomalies across India is contested, its relationship with \enso/ (ENSO) has weakened since the early 20th century.
% The JJAS rainfall regression patterns for ENSO and IOD change in compensating ways from the pre-satellite to satellite eras, such that a combined index is more stable.
\end{abstract}

% 200 word limit; required for GRL
\section*{Plain Language Summary}
The Indian summer monsoon generates copious rainfall each June through September, but more in some years than others, and more in some Indian sub-regions than others.  We use observation-based datasets spanning 1901-2020 to reconcile past disagreements about these fluctuations.  There has long been concern that the rainfall rate averaged over the whole summer and whole of India---the All-India Rainfall Index---doesn't necessarily track with the spatial extent of wet or dry anomalies within India which is more relevant for many societal purposes, but we show that the two measures vary closely with one another.  There has been disagreement about how the all-India average relates to rainfall in fixed sub-regions of India, and we show using high-resolution rainfall data that the issue stems in part from the use of coarse datasets in the past.  Finally, we reconcile disagreements about whether or not the Indian Ocean Dipole (IOD) teleconnection influences all-India rainfall by showing that, when the El Ni\~no-Southern Oscillation signal that influences both the IOD and the monsoon rains is removed, the IOD significantly influences the all-India average and with a distinct sub-India spatial pattern.

\section{Introduction}
\label{sec:intro}

% Indian summer monsoon rainfall varies on temporal scales from hourly \cite{moron_storm_2021} to centennial \cite{sinha_leading_2011} and spatial scales from kilometers or less to the whole country.  This rich spectrum of variability in Earth's strongest monsoon circulation \cite{nie_observational_2010} profoundly influences the lives of India's more than billion citizens and Earth's general circulation.  As such, accurate predictions of Indian summer monsoon rainfall variability have been sought since at least the 19th century \cite[\eg/ reviews by][]{gadgil_indian_2003,gadgil_monsoon_2018}.  Prediction efforts often focus on arguably the simplest bulk measure of the summer monsoon, the all-India rainfall average, or AIR, averaged over the summer monsoon season of June-July-August-September (JJAS).  Societally, JJAS AIRI is known to influence all-India grain yields \cite{mooley_annual_1981} and is well correlated with interannual deviations in Indian agricultural yields and gross domestic product \cite{gadgil_indian_2006}.  Physically, it is intuitive and a direct measure of the total diabatic heating generated over the subcontinent by the summer monsoon (though importantly this neglects the rainfall over neighboring countries and oceanic points).

Rainfall averaged across India and over June-July-August-September (JJAS)---the All-India Rainfall index (AIRI)---is a widely used, physically intuitive, and societally relevant \cite{mooley_annual_1981,parthasarathy_regression_1988,gadgil_climate_1995,gadgil_indian_2006} bulk indicator of the summer monsoon.  But its use is controversial, largely owing to summer monsoon rainfall's pronounced sub-India and sub-seasonal heterogeneity.  Summer rainfall is high within a broad band of northern-central India, higher still within a narrow band in the southwest between the Arabian Sea and the Western Ghats mountains, and much lower in southeastern India between these two bands. And throughout the subcontinent JJAS rainfall variability is much higher on daily than interannual timescales \cite{krishnamurthy_intraseasonal_2000}.  In this study we address three long-running controversies regarding AIRI's interpretation in light of this spatiotemporal variability, attempting to synthesize and reconcile past arguments made with differing and often short and/or coarse datasets by using state-of-the-art, 120-year datasets of Indian rainfall at high resolution and of sea surface temperature (SST).

First is the relationship between AIRI and the overall spatiotemporal extent of wetting or drying across India.  Spatially, in most summers there are parts of India that experience drought and others excess rainfall, a heterogeneity in summer-mean rainfall anomalies that led \citeA{gadgil_climate_1995} to describe AIRI as ``not very meaningful.''  Indeed, for rain-fed agriculture quite plausibly the spatial extent of drought relative to local rainfall normals is more relevant than the average rainfall anomaly in \mmday/ across India \cite{parthasarathy_homogeneous_1993}, leading to past attempts to define bulk indices in terms of the fraction of the Indian surface area experiencing rainfall anomalies exceeding specified thresholds \cite{mooley_annual_1981,parthasarathy_droughtsfloods_1987}.
% ---but subsequently demonstrates that AIRI is well correlated with interannual deviations in all-India annual rice and total foodgrain production.
% On similar grounds, \citeA{mooley_annual_1981} generate timeseries of the percentage of Indian land experiencing a given severity of rainfall excess or deficit, which they consider more relevant to agricultural outcomes than the rainfall integral measure that is AIRI.
% \citeA{parthasarathy_droughtsfloods_1987} present timeseries of the fraction of India experiencing drought or flood (based on IMD meteorological subdivisions with a JJAS rainfall anomaly exceeding 25\% of the mean) but do not explicitly compare them to AIRI.
Temporally, a recent series of studies \cite{moron_impact_2012,moron_spatial_2017,robertson_multi-scale_2019} employ a useful decomposition of JJAS-mean rainfall at any given point into the product of the frequency of rainy days and the mean rainfall intensity on rainy days, arguing that rainy day frequency tends to be more spatially coherent across India and thus potentially predictable.  We will make use of this decomposition to construct and compare alternative indices to AIRI.

Second is the relationship between AIRI and rainfall anomalies over geographically fixed sub-India regions.  The literature on dividing India into so-called ``coherent'' or ``homogeneous'' sub-regions---\ie/ over which rainfall interannual variability tends to be similar---dates to the 19th century (see \citeA{shukla_interannual_1987,parthasarathy_homogeneous_1993,gadgil_coherent_1993} and references therein).  Subseasonally, it is well established that the predominant variability mode---the so-called active-break cycle---comprises a tripole: in active periods, rainfall is enhanced over the Western Ghats (henceforth WG) and Central Monsoon Zone (henceforth CMZ) bands but suppressed over southeastern India (henceforth SEI) \cite{rajeevan_active_2010}.  And because the CMZ and WG anomalies are larger than the SEI anomalies in absolute terms, AIRI is enhanced (the same results hold with signs reversed for break periods).
% This CMZ-SEI-WG tripole pattern emerges in numerous subseasonal contexts such as: daily snapshots of satellite cloud cover imagery \cite{sikka_maximum_1980}; composites based on the presence of monsoon low pressure systems \cite{krishnamurthy_composite_2010}; cluster analysis of \(\sim\)monthly rainfall variability \cite{moron_spatial_2017}; composites on the phase of the Madden-Julian Oscillation \cite{pai_impact_2011}; composites based on intraseasonal active/break phases of monsoon rainfall \cite{krishnamurthy_intraseasonal_2007,krishnamurthy_seasonal_2008,rajeevan_active_2010}; and the first empirical orthogonal function (EOF) computed from daily JJAS rainfall standardized anomalies \cite{krishnamurthy_intraseasonal_2000}.
Interannually, the predominant mode is instead a quasi-uniform wetting or drying across the subcontinent that strongly projects onto AIRI \cite{shukla_interannual_1987,krishnamurthy_intraseasonal_2000,krishnamurthy_intraseasonal_2007,krishnamurthy_seasonal_2008,straus_preferred_2007,mishra_prominent_2012,moron_impact_2012,moron_spatial_2017}.  A second seasonal-mean variability mode has been identified \cite{krishnamurthy_seasonal_2008} but characterized by \citeA{mishra_prominent_2012} as a north-south dipole rather than the CMZ-SEI-WG tripole, with generally same-signed anomalies across peninsular India including both the WG and SEI, and opposite-signed conditions to the north including the CMZ.  This would seem consistent with the argument put forth separately that rainfall anomalies in the CMZ and WG are weakly correlated, and that therefore AIRI effectively amounts to the superposition of two quasi-independent processes \cite{vecchi_interannual_2004}.  However, we will argue that the \(\geq\)1\degr{}\(\times\)1\degr{} rainfall datasets used regarding both points lead to erroneous or incomplete conclusions.

Third is the relationship between both AIRI and sub-India rainfall anomalies to teleconnection modes emanating from the tropical Pacific and Indian Oceans. % (neglecting for brevity roles potentially played by the tropical Atlantic, the extratropical oceans, and land and cryosphere processes).
The link between the Indian summer monsoon and the \enso/ (ENSO) is well established, with drought more likely in \elnino/ summers and excess rain in \lanina/ summers \cite{pant_aspects_1981,rasmusson_relationship_1983}, both at the all-India scale and most individual points (again excluding the far northeast).  Nevertheless, ENSO indices have always left the majority of AIRI variance unexplained \cite{surendran_prediction_2015}, even before a weakening of the ENSO-AIRI lag-zero correlation in recent decades compared to the mid 20th century \cite{kumar_weakening_1999,ashok_indian_2019}.  The Indian Ocean Dipole (IOD)---the zonally oriented oscillatory mode spanning the equatorial Indian ocean \cite{saji_dipole_1999,webster_coupled_1999}---has been argued to modulate ENSO's influence on AIRI \cite{ashok_impact_2001,ashok_individual_2004}, with the IOD-AIRI correlation relatively high in decades when the ENSO-AIRI correlation is relatively low and vice versa.  While it has been argued that the IOD is in fact largely controlled by ENSO \cite{krishnamurthy_variability_2003,stuecker_revisiting_2017}, \citeA{ashok_impacts_2007} argue that the IOD spatial imprint on Indian summer monsoon rainfall is much more heterogeneous, concentrated  roughly over the CMZ region.
% The direction of causality is not clear, with the surface winds related to monsoon circulation variability thought to influence the IOD \cite{webster_coupled_1999,saji_dipole_1999}.
Meanwhile, other studies focused on the atmospheric counterpart to the IOD, the Equatorial Indian Ocean Oscillation (EQUINOO), have argued that the correlation between the IOD and AIRI is effectively zero \cite{gadgil_extremes_2004,ihara_indian_2007,surendran_prediction_2015}.  We will attempt to reconcile these arguments by examining the spatial imprint of the IOD and ENSO on rainfall amount, rainy-day frequency, and mean rainfall intensity (c.f. \citeA{moron_spatial_2017}) and by isolating the IOD and ENSO signals from one another (c.f. \cite{ashok_impacts_2007}).

% EQUINOO can be thought of as the atmospheric counterpart to IOD, albeit with weaker atmosphere-ocean coupling between EQUINOO and the IOD compared to ENSO \cite{gadgil_droughts_2003,gadgil_extremes_2004}.  In EQUINOO's positive phase, convection is enhanced in the western equatorial Indian Ocean (and over the subcontinent) relative to the eastern equatorial Indian Ocean.  Several studies have found the JJAS lag-zero correlation between EQUINOO and AIRI to be considerably higher than between IOD and AIRI \cite{gadgil_extremes_2004,ihara_indian_2007,surendran_prediction_2015}.

% JJAS rainfall averaged over the CMZ is highly correlated with the all-India average,  \cite{gadgil_indian_2003,rajeevan_active_2010}.

% Understanding the roles of processes other than ENSO (such as the IOD and EQUINOO) in Indian summer monsoon rainfall is complicated by additional ENSO-related factors including nonlinear monsoon rainfall dependence on ENSO, \elnino/ flavors, the SST zonal gradient across the eastern equatorial Pacific, tropical Indian Ocean warmth linked to the preceding winter's ENSO, and combinations thereof.

After describing the datasets and methods used (Section~\ref{sec:methods}), we argue the following (Section~\ref{sec:results}): that AIRI is an excellent indicator of the spatial extent of wet anomalies across the subcontinent; that the second leading mode of interannual variability is the familiar CMZ-SEI-WG tripole; that rainfall anomalies within the CMZ and WG regions are well correlated; that while ENSO strongly imprints on the leading monopole pattern and thus AIRI, the IOD imprints on the tripole pattern and thus weakly on AIRI; and finally that when the ENSO signal is linearly removed the IOD influence becomes more positive at most points such that the ENSO-residual IOD and AIRI timeseries are significantly correlated.  We then conclude with summary and discussion (Section~\ref{sec:conc}).

\section{Methods}
\label{sec:methods}

For rainfall, we use the Indian Meteorological Department (IMD) daily, gridded 0.25\(\times\)0.25\degr{} dataset spanning 1901-2020, which is derived from a dense but time-varying network of rain gauges across India comprising 6955 in total and an average of around 2,600 operational in each year \cite{pai_development_2014}.   Daily values from 1 June to 30 September are averaged to form a JJAS seasonal mean for each year.  We have compared the IMD dataset with the TRMM 3B42v7 \cite{huffman_trmm_2007} daily dataset over the period 1998-2014.  While on any given day the differences calculated from the two datasets over India can be substantial (not shown), the JJAS AIRI timeseries between the two datasets are very highly correlated (\(r=0.94\)).

As common in studies of the summer monsoon \cite{parthasarathy_regression_1988,krishnamurthy_intraseasonal_2000,vecchi_interannual_2004}, we exclude points in far north and northeast India, in our case by applying the ``monsoonal India'' mask of \citeA{gadgil_new_2019}.  In far northern India the terrain makes rain gauges less reliable and representative, and the vicinity to the Himalayas and extratropical influences make interannual rainfall variability not clearly related to the summer monsoon circulation.  For Northeast India there appear to be spurious data points in a cluster of grid cells on the border of Bangladesh, with the JJAS mean and variance jumping in 1971 to implausible values over the remainder of the record (not shown).  Nonetheless, the mean JJAS rainfall restricted by the monsoon mask is very highly correlated with the all-India rainfall average computed with these spurious points included, (\(r=0.95\)).  For simplicity henceforth we will refer to the monsoonal-India average as AIRI.

We use standard SST-based indices of ENSO and the IOD---the NINO3.4 and Dipole Mode Index (DMI; \citeA{saji_dipole_1999}) respectively---computed with monthly SSTs from the NOAA Extended Reconstruction SST \cite{huang_extended_2015} dataset (ERSST) version 5, 1901-2020.
% , and daily SSTs from the daily NOAA OISST dataset \cite{huang_improvements_2021}, 1982-2020.
% We use 10-m zonal wind and outgoing longwave radiation (OLR) from the CERA-20C, ensemble-based, extended reanalysis dataset \cite{laloyaux_cera-20c_2018} that only assimilates surface observations and spans 1901-2010, and from the ERA5 reanalysis dataset  \cite{hersbach_era5_2020}, 1950-2020 (note that 1950-1978 uses the provisional ``extended'' portion of the ERA5 dataset), with fields from the two reanalyses merged into a 1901-2020 timeseries as follows.  For 1901-1949, monthly, ensemble-mean fields are taken from CERA-20C.  For 1950-2020, hourly winds from ERA5 are averaged into daily and JJAS means.  We concatenate the two timeseries (\ie/ CERA-20C for 1901-1949 and ERA5 for 1950-2020), subtract off the linear trend over 1901-2020, and finally normalize by the standard deviation.
NINO3.4 is defined as the average SST anomaly spanning 120-170\degr{}W, 5\degr{}S-5\degr{}N, and DMI is defined as the difference between the SST anomalies averaged over a western equatorial Indian Ocean box (90-110\degr{}E, 10\degr{}S-0\degr{}) and over an eastern equatorial Indian Ocean box (50-70\degr{}E, 10\degr{}S-10\degr{}N).  Anomalies are defined as the deviation at each gridpoint from the local climatology over 1901-2020 for that month, with June through September anomalies then averaged to yield JJAS values for each index.
To focus on interannual variability, for all analyses we first subtract off the linear trend computed from simple least squares regression of the 1901-2020 timeseries of JJAS-mean values.  Because 1901-2020 trends are weak at the vast majority of gridpoints as well as for the teleconnection indices (not shown), all major results we present are insensitive to this choice.
% Daily anomalies are computed by first computing a daily climatology as the average across years on that calendar day and then taking the difference between that day's full field and this daily climatology.
% Where relevant, we further remove the seasonal-mean anomaly by then subtracting off that year's JJAS-mean anomaly.
% The daily climatology is then approximated by the first two annual harmonics to smooth out sampling noise.
When decomposing JJAS-mean rainfall into the product of the number of rainy days times the mean rain intensity on rainy days, we use a 1~\mmday/ threshold to define rainy days, but results are very similar if 0~\mmday/ is used instead (not shown).  For all detrended JJAS timeseries analyzed the interannual autocorrelations are sufficiently weak and the distributions are sufficiently normal (not shown) that a simple t-test yields a reasonable measure of statistical significance; in particular, using a conservative estimate of 100 degrees of freedom yields significance at the 5\% level for correlations with \(|r|>0.195\).  Though it should be noted that for spatially gridded data such a point-by-point approach likely overstates significance to some degree due to spatial autocorrelations \cite{wilks_stippling_2016}.

\section{Results}
\label{sec:results}
% \section{All-India and sub-India JJAS rainfall variations and the ENSO teleconnection}
% \label{sec:rain}

\subsection{AIRI relationship to spatiotemporal extent of rainfall anomalies}
\begin{figure}[t]
  \centering\noindent
  \includegraphics[width=\textwidth]{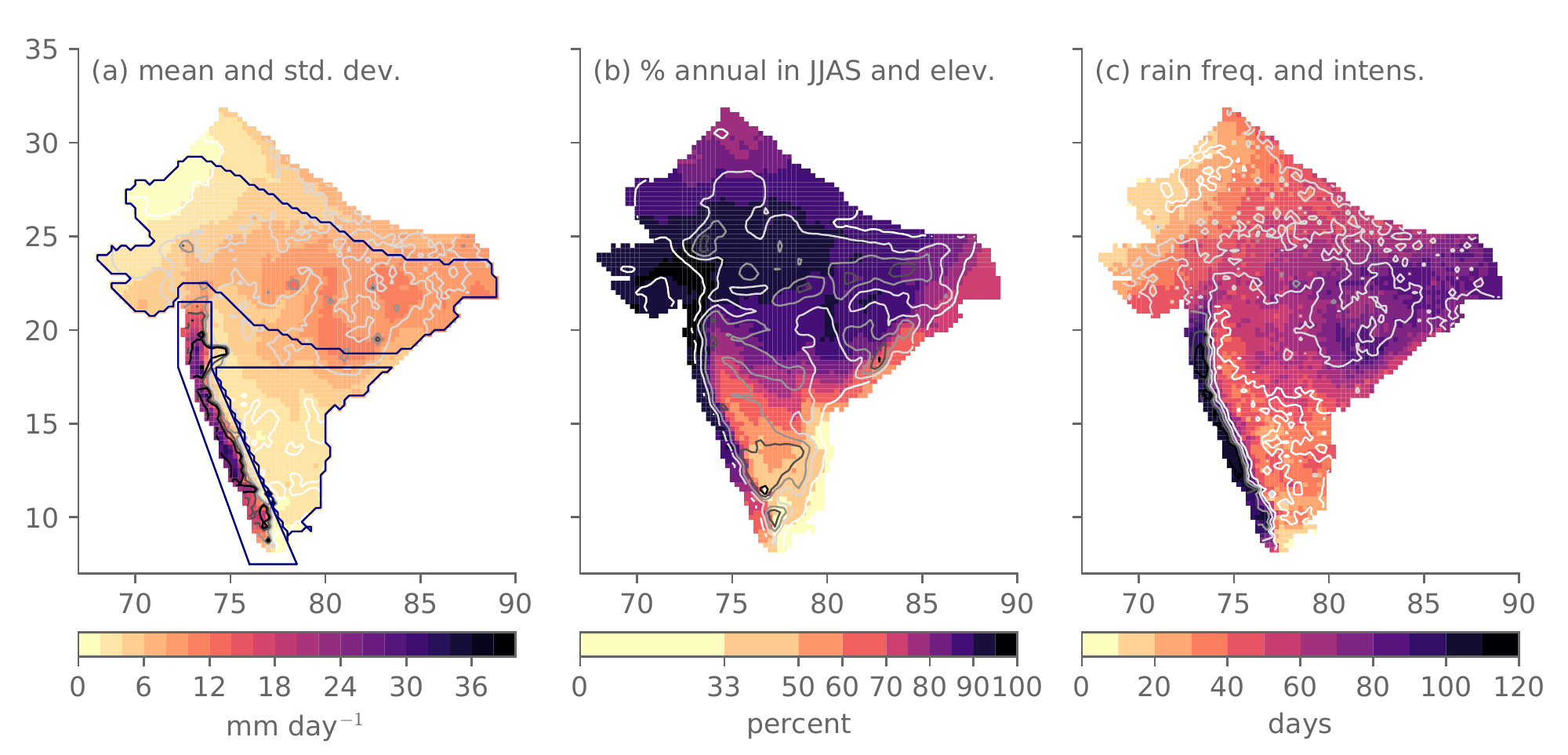}\\
  \caption{(a) Climatological JJAS rainfall mean in color shading and standard deviation in contours in the IMD 0.25\(\times\)0.25\degr{} gridded dataset, 1901-2020.  Standard deviation contours from white to black are 1, 2, 3, 4, and 5 \mmday/ (maximum standard deviation is 14.2~\mmday/).  (b) Climatological fraction of annual-mean rainfall that occurs during JJAS in color shading and surface elevation in contours.  Elevation contours from white to black are 100, 300, 500, 700, and 900~m (maximum elevation is 1028~m).  (c) Climatological mean number of rainy days in JJAS in color shading and mean rainfall intensity on rainy days in contours.  Mean rainfall intensity contours from white to black are 10, 15, 20, 25, and 30~\mmday/ (maximum mean rainfall intensity is 44.8~\mmday/.  Blue outlines in panel a and in subsequent figures show the borders of the three sub-India regions we analyze in detail, namely the Central Monsoon Zone (CMZ) in the north, Western Ghats (WG) in the southwest, and Southeastern India (SEI) in the southeast.}
  \label{fig:mean-stdev}
\end{figure}

Fig.~\ref{fig:mean-stdev}(a) shows the climatological JJAS rainfall and its interannual standard deviation for each gridpoint within monsoonal India.  The standard deviation largely scales with the local climatological mean, and both are heterogeneous---high within CMZ, highest within WG, and much lower in SEI.  Fig.~\ref{fig:mean-stdev}(b) shows the fraction of annual-mean rainfall that occurs in JJAS as well as surface elevation contours.  The JJAS rainfall fraction exceeds 95\% in parts of the far western CMZ, exceeds 80\% over most of the WG and CMZ, and does not exceed 50\% in most of SEI, where in the far southeast the rainy season occurs during the northeast monsoon in boreal autumn \cite{ramesh_globally_2021}.  The elevation contours highlight how southwesterly low-level summer monsoon flow largely perpendicular to the Western Ghats mountain range axis results in a concentrated rainfall band on the windward side and a much broader and drier regime on the leeward side over the Deccan Plateau \cite{francis_intense_2006,flynn_mesoscale_2017,hunt_modes_2021}.
Fig.~\ref{fig:mean-stdev}(c) shows the climatological number of rainy days in JJAS and the mean rainfall rate on those rainy dates.  Rain falls nearly every summer day within most of WG, most days within most of the CMZ, and less frequently within most of SEI.  Mean rainfall intensity is highest within WG directly along the coast, with a sharp gradient moving inland up the mountain slope, with smoother spatial variations elsewhere.

% The CMZ and WG surface areas are 52.1 and 9.3\% respectively of the monsoonal India surface area, and climatologically they harbor 52.8 and 19.8\% of the rain that falls in monsoonal India in JJAS.

Given this heterogeneity, in principle AIRI need not be a useful indicator of the spatial extent of wet or dry anomalies: an average across regions with high variance and regions with low variance will depend primarily on the former and weakly on the latter.  We have therefore constructed six alternative all-India measures of JJAS rainfall variability: first, the average standardized rainfall anomaly (\ie/ the raw anomaly divided by the standard deviation, rather than just the raw anomaly as for AIRI); second, the number of gridpoints in which the local JJAS rainfall anomaly is positive; third, the average across all gridpoints of the number of rainy days; fourth, the average across all gridpoints of the rainfall rate on rainy days; fifth, the mean rainfall anomaly restricting to those gridpoints for which the rainfall anomaly is positive; and sixth, the mean rainfall anomaly restricting to those gridpoints for which the rainfall anomaly is negative.  The first, third, and fourth variants were introduced by \citeA{moron_spatial_2017}.

\begin{figure}[t]
  \centering\noindent
  \includegraphics[width=\textwidth]{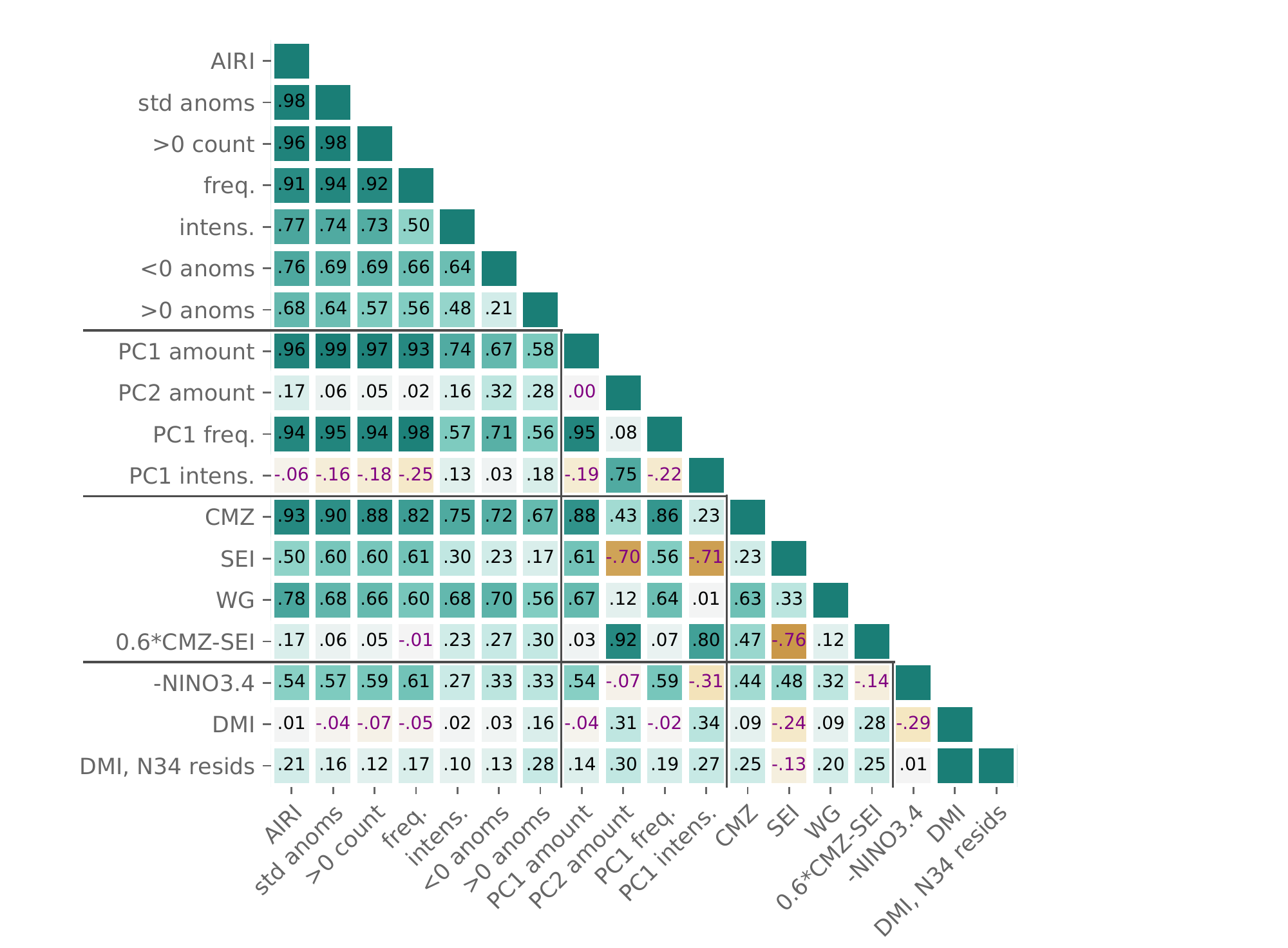}\\
  \caption{Cross-correlation matrix showing the correlation coefficients (unitless) among AIRI, six variants thereof as defined in the text, four PC timeseries as described in the text, regional rainfall averages over each of the CMZ, SEI, and WG regions and a linear combination of CMZ and SEI as described in the text, and the NINO3.4 (with sign reversed) and DMI timeseries, all for JJAS and over the 1901-2020 period.  % The color of each box corresponds to the sign and magnitude of that correlation coefficient.
For the final row only, labeled ``DMI, N34 resids'', both the DMI timeseries and all target timeseries are the residuals after linearly removing the NINO3.4 signal.  Horizontal and vertical lines separate the timeseries into categories: AIRI and variants thereof, PC timeseries, sub-region averages, and teleconnection indices, respectively from top to bottom and left to right.}
  \label{fig:cross-corrs}
\end{figure}

Fig.~\ref{fig:cross-corrs} provides a cross-correlation matrix of AIRI and these six alternative measures (along with numerous additional fields below the first horizontal line to be discussed below).  Among AIRI and the first three variants---which characterize the spatial or temporal extent of wet anomalies---all correlation coefficients exceed 0.9.  Among AIRI and the last three variants, which characterize the intensity of wet or dry anomalies, no correlation coefficient exceeds 0.77.  Thus, AIRI depends more on the spatiotemporal extent of rainfall across India and JJAS than on the severity of rainfall anomalies when and where they occur.  This agrees with \citeA{moron_spatial_2017}, who also note that this contrasts with interannual variability at the gridpoint scale: for most gridpoints of monsoonal India the JJAS rainfall amount depends more on the intensity of rain events than the number of rainy days.

We therefore conclude that AIRI is an excellent indicator of the overall spatial extent of wet or dry anomalies, as well as of the average anomaly in rainy day occurrence across India.  Insofar as the latter two measures are more directly relevant to agricultural and thus economic outcomes, this helps explain the high correlations found by previous studies of AIRI itself with agricultural yields and gross domestic product on interannual timescales \cite{mooley_annual_1981,parthasarathy_regression_1988,gadgil_climate_1995,gadgil_indian_2006}.
% At the same time, these indices are agnostic as to where within India or when within JJAS there will be deficient or excess rainfall, and so we now turn to the spatial imprint of interannual variability modes in summer monsoon rainfall.

\subsection{Rainfall relationships among fixed sub-India regions}
\begin{figure*}[t]
  \centering\noindent
  \includegraphics[width=0.7\textwidth]{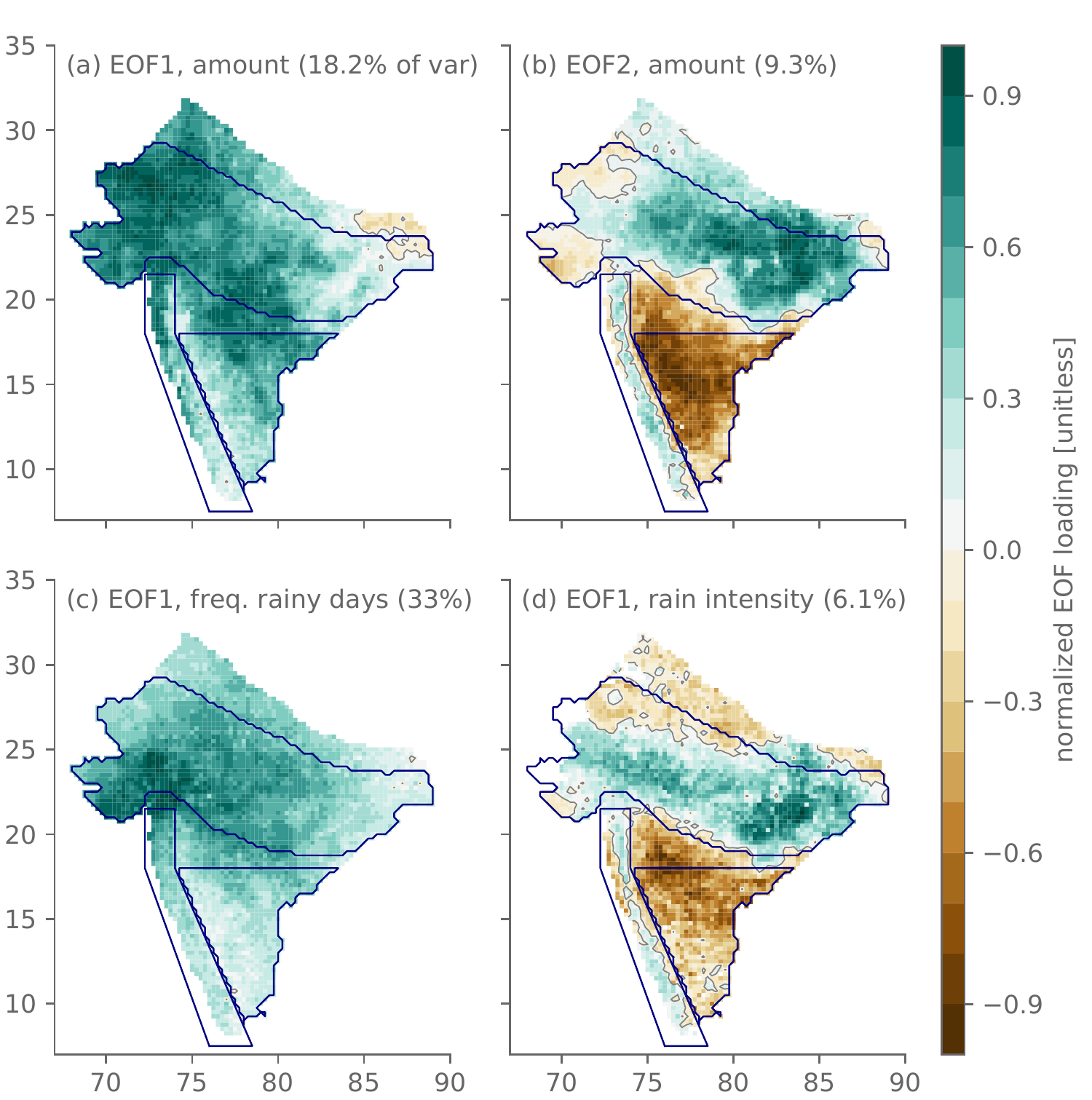}\\
  \caption{(a) First and (b) second EOF of JJAS rainfall amount, and first EOF of (c) rainy day frequency, and (d) rainfall intensity, each with the percentage of variance explained by that EOF printed in the top-left.    In each panel, values are normalized by that EOF's maximal value, and solid gray contours signify that EOF's zero contour.}
  \label{fig:eofs}
\end{figure*}

We have repeated and extended the empirical orthogonal function (EOF) analysis of \citeA{moron_spatial_2017} for JJAS rainfall amount, rainy day frequency, and mean rainfall intensity, by including more recent years and including the second rainfall amount EOF \cf/ \citeA{mishra_prominent_2012}.  Fig.~\ref{fig:eofs} shows the first EOF for each and the second EOF for rainfall amount, and Fig.~\ref{fig:cross-corrs} reports correlation coefficients among the corresponding principal component (PC) timeseries and AIRI.  The first EOF of the rainfall amount accounts for 18.2\% of the total variance and consists of same-signed loadings at the vast majority of gridpoints \cite{krishnamurthy_intraseasonal_2000}.  This resembles the first EOF of rainy day frequency, which explains 32.4\% of the rainy day frequency variance and exhibits an even more homogeneous pattern, though sharing the two local minima located east of the WG mountains axis and in the far east.  The PC timeseries corresponding to these amount and frequency first EOFs are each highly correlated with AIRI (\ilcorr{0.96} and 0.93, respectively).

Both the first EOF of mean rainfall intensity (7.2\% of variance) and the second EOF of rain amount (9.3\% of variance) comprise a CMZ-SEI-WG tripole pattern.  The corresponding PCs are highly correlated with one another (\ilcorr{0.85}) but weakly with AIRI.  Unlike for rainfall amount and rainy day frequency, the first EOF for mean rainfall intensity is not well separated in variance explained from the remaining EOFs (the second EOF, not shown, accounts for 5.5\%).  But its physical relevance is supported by the correspondence of its pattern to the rainfall EOF2 and to the IOD teleconnection results we will present below.
% The rainfall amount EOF2 \sah{IS/ISN'T} well separated from EOF3.

% Summarizing, then, to first order AIRI is controlled by the extent of wet anomalies over monsoonal India, or equivalently by the anomalous number of rainy days averaged across monsoonal India.  The leading mode of JJAS rainfall variability strongly is nearly unipolar, quasi-homogeneous, is primarily determined by the corresponding mode controlling the number of rainy days, and strongly influences AIRI.  The second mode of JJAS rainfall variability strongly is a CMZ-SEI-WG tripole, is primarily determined by the corresponding mode controlling the number of rainy days, and weakly influences AIRI.

% We have computed composites over the eight driest (1918, 2002, 1987, 1972, 1905, 1965, 1974, 1979) and, separately, eight wettest (1961, 1917, 1994, 1959, 1975, 1933, 2019, 1983) years in terms of the detrended JJAS AIRI anomaly (not shown).  The spatial structures of the two composites are similar (pattern correlation \ilcorr{-0.72}), and the difference between them closely resembles closely resembles the EOF1 computed for raw rainfall anomalies.

We compute averages over each of the CMZ, WG, and SEI sub-regions using border definitions as follows (and are shown as blue contours in Fig.~\ref{fig:mean-stdev}(a)).  The Central Monsoon Zone (CMZ) boundaries follow \citeA{gadgil_new_2019}.  We define the non-coastal borders of WG to align with the sharp rainfall gradient separating the coastal and inland peninsular regimes,
% \footnote{In terms of IMD meteorological subdivisions, the WG region roughly corresponds to, but extends further inland than, the union of three existing subdivisions, Konkan and Goa, Coastal Karnataka, and Kerala \cite{kelkar_meteorological_2020}.}
and define (SEI) to be all gridpoints west thereof and south of 18\degr{}N.
% Fig.~\ref{fig:corr-pointwise} shows the correlation between JJAS-mean rainfall at each gridpoint with one selected gridpoint for each region.
Pointwise correlation maps (not shown) indicate that rainfall in points within the CMZ tends to vary positively both with other CMZ points and with points within WG, but weakly or negatively with those in SEI; the region-mean JJAS rainfall correlations are \ilcorr{0.23}, 0.63, and 0.33 respectively for CMZ-SEI, CMZ-WG, and SEI-WG.

Using the CMZ, SEI, and WG regional averages, we have constructed one, two, and three-region linear regression models for AIRI and for the tripole mode as indicated by the rain amount PC2 (results are similar if the intensity PC1 is used instead; not shown).  For AIRI, at \ilcorr{0.93} the CMZ alone explains most of the variance as shown by \citeA{gadgil_new_2019}, leaving little additional skill to be gained by including either of the other regions.  For the tripole mode, by contrast, the highest-magnitude single-region correlation is with SEI at \ilcorr{-0.70}.  The CMZ-SEI two-region model, whose relative weights are 0.6 and -1 respectively and whose timeseries is included in Fig.~\ref{fig:cross-corrs} along with the three individual region averages, yields \ilcorr{0.92} with the tripole mode.  This is effectively unchanged by including the WG as well.

% \begin{table}[t]
% \caption{Regression models for AIRI and PC2 of the JJAS rainfall amount over monsoonal India in terms of the raw rainfall averages over the CMZ, SEI, and WG sub-India regions.  The first row indicates which of the three regions is included.  For each index, the first row shows the Pearson correlation coefficient of the regression model with the index, and the second row shows for the two- and three-region models the weights for each region, normalized by the largest weight.}\label{table:regr-models}
% \begin{center}
% \begin{tabular}{lrrrrrrr}
% \hline\hline
% ***SAH NOTE: the SEI values here use an old boundary; if table is to be included would need to replace the SEI values with the correct ones.***
%                     & CMZ  & SEI     & WG   & CMZ, SEI   & CMZ, WG    & WG, SEI    & CMZ, SEI, WG \\
% \hline
% AIRI, \(r\)         & 0.93 &    0.58 & 0.78 & 0.98       & 0.96       & 0.83       & 0.99 \\
% AIRI, weights       &      &         &      & 1, 0.48    & 1, 0.19    & 0.75, 1    & 1, 0.46, 0.15\\
% \hline
% Amount PC2, \(r\)   & 0.43 & $-$0.64 & 0.12 & 0.92       & 0.47       & 0.75       & 0.92 \\
% Amount PC2, weights &      &         &      & $-$0.44, 1 & 1, $-$0.20 & $-$0.13, 1 & -0.41, 1, 0.02\\

% \hline
% \end{tabular}
% \end{center}
% \end{table}

Thus, despite WG being home to the highest JJAS-mean rainfall variances throughout monsoonal India, it adds little predictive power to CMZ and SEI for these bulk measures, due to it being well correlated with CMZ (\ilcorr{0.63}).  This is considerably higher than two similar-but-different regions (\(r\approx0.2\)) reported by \citeA{vecchi_interannual_2004}.  That study's WG region (borders overlain in Fig.~\ref{fig:nino34-regress-maps}d), based on coarse 2.5\(\times\)2.5\degr{} gridded CMAP rainfall data, spans into peninsular India and thus includes points for which rainfall anomalies are anti-correlated with those along the coast.  The correspondence between WG and CMZ rainfall anomalies has also been established through cluster analysis of seasonal-mean rainfall anomalies by \citeA{moron_spatial_2017} using 0.25\degr{}\(\times\)0.25\degr{} data.
% Using the \citeA{vecchi_interannual_2004} region definitions applied to the IMD \(0.25^\circ\times0.25^\circ\) dataset 1901-2020 we use, the cross-region correlation is low (not shown).  A \(2.5^\circ\times2.5^\circ\) grid is simply too coarse to resolve the\(\sim1^\circ\)-wide coastal rainfall band nor the sharp gradient between it and the rain-shadow just beyond.

We conclude that the sharp rainfall gradients in the southwest---which have dictated the border definitions of subdivisions of the Indian Meteorological Department station network since the 19th century \cite{kelkar_meteorological_2020}---are insufficiently resolved in gridded datasets with resolutions of 1\degr{} or more.  With higher resolution the second interannual variability mode after the quasi-uniform first mode is a CMZ-SEI-WG tripole.  A semblance of the tripole is apparent in Fig.~1 of \citeA{mishra_prominent_2012} based on 0.5\(\times\)0.5\degr{} data, but it is described in their text as a north-south dipole.  As such, the CMZ and WG mean rainfall anomalies cannot be considered quasi-independent \cite{vecchi_interannual_2004}.

\subsection{ENSO and IOD teleconnections}
\label{sec:telecon}
Fig.~\ref{fig:cross-corrs} includes the -1\(\times\)NINO3.4 and DMI timeseries, the sign of NINO3.4 flipped so that positive values correspond to positive AIRI anomalies.  The correlation between -1\(\times\)NINO3.4 and AIRI is \ilcorr{0.54}, between \(-1\times\)NINO3.4 and DMI is \ilcorr{-0.29}, and between DMI and AIRI nearly zero, \ilcorr{0.01}, all broadly consistent with past studies.  Using partial correlation analysis to remove the linear NINO3.4 signal from both rainfall and the DMI timeseries, the resulting DMI and AIRI residual timeseries become significantly positively correlated, \ilcorr{0.21}.  Conversely, removing the DMI signal hardly affects the ENSO-AIRI relationship (\ilcorr{0.56}), and mechanisms through which ENSO influences the IOD are well documented \cite{stuecker_revisiting_2017}.  We therefore now analyze the spatial imprints of ENSO and the IOD via the frequency-intensity decomposition, in the case of the IOD focusing on the NINO3.4 residuals (\cf/ last row of Fig.~\ref{fig:cross-corrs}).

\begin{figure*}[t]
  \centering\noindent
  \includegraphics[width=\textwidth]{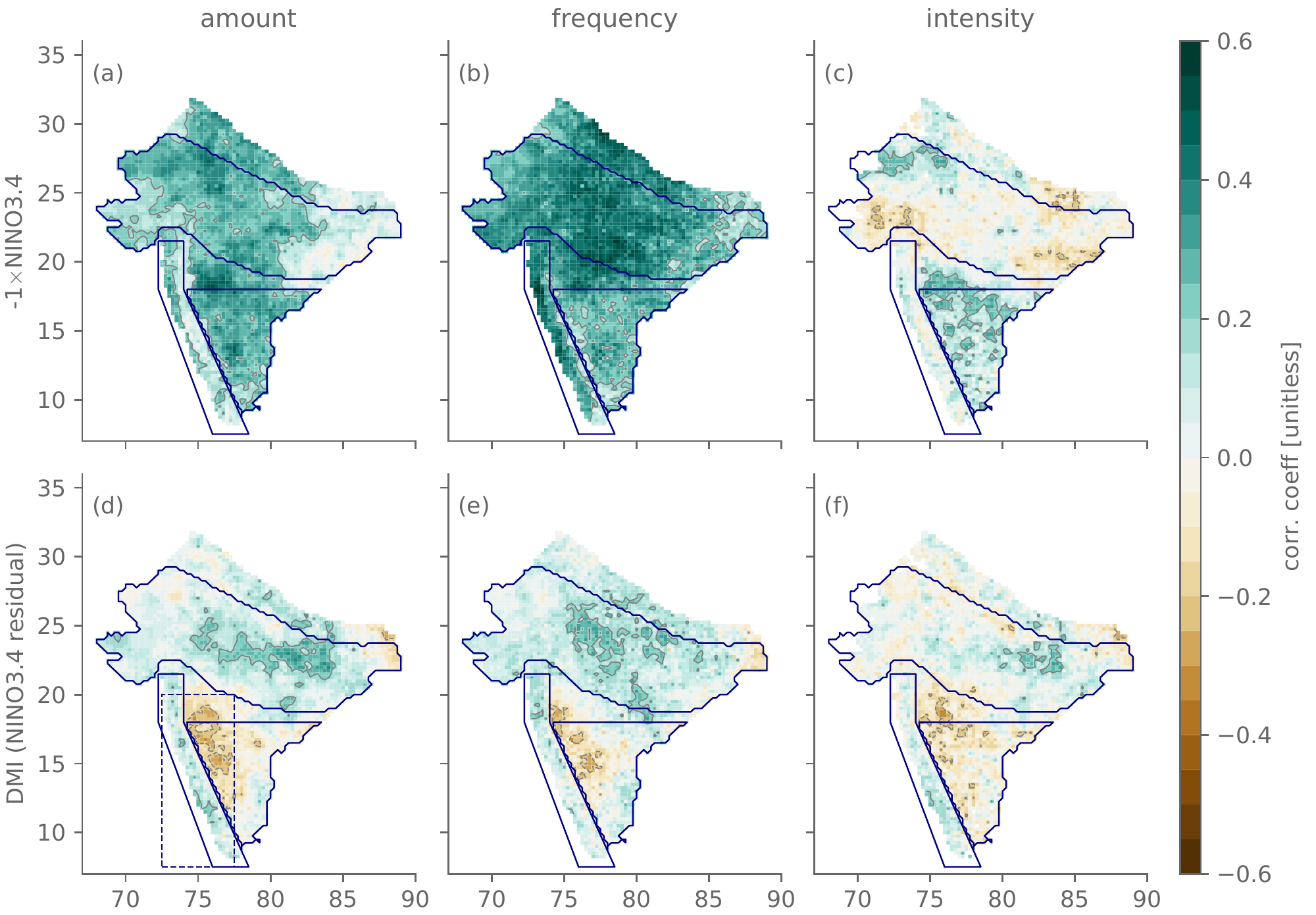}
  \caption{Correlation coefficient between JJAS -1\(\times\)NINO3.4 and gridpoint local (a) rainfall amount, (b) frequency of rainy days, and (c) mean rainfall intensity.  Overlain in thin gray are the -0.2 and +0.2 contours, which give a reasonable estimate of statistical significance as described in Section~\ref{sec:methods}.  Dashed box in (d) shows the WG region border used by \citeA{vecchi_interannual_2004}.}
  \label{fig:nino34-regress-maps}
\end{figure*}

Fig.~\ref{fig:nino34-regress-maps}(a-c) shows the correlation of each of JJAS rainfall amount, rainy day frequency, and mean rainfall intensity on \(-1\times\)NINO3.4.  For rainfall amount, the correlation is positive at most gridpoints, most positive within SEI and the central portion of CMZ, and least positive (including some slightly negative values) in the far east of the CMZ and along the Western Ghats mountain axis.
% This homogeneity results from compensating factors: large variance but modest correlations in the CMZ \vs/ small variance but larger correlations in SEI.  A local minimum in the correlation runs along the axis of the Western Ghats mountain range, which explains the modest regression there compared to the coastal points just to the west.
The rainy day frequency correlation is more homogeneously positive and with higher magnitudes in most areas, especially along the Arabian Sea coast within WG and the central and western CMZ.  The mean rainfall intensity correlation is noisier and has generally weaker magnitudes, with a considerable fraction of gridpoints below the approximate statistical significance threshold of \(|r|\gtrsim0.2\) described in Section~\ref{sec:methods}.  It takes the reverse sign over an appreciable fraction of gridpoints including much of the CMZ, and of the three sub-regions it is most consistently positive over SEI.  These three spatial patterns are broadly comparable to the corresponding first EOFs for JJAS rainfall amount (mostly same-signed), frequency (more homogeneous than amount), and intensity (heterogeneous with a CMZ-SEI-WG tripole; the weaker statistical significance notwithstanding).

% Because AIRI is imperfectly correlated with NINO3.4, the resulting residual AIRI timeseries is well correlated with the original AIRI timeseries (\ilcorr{0.84}).
% We have performed EOF analysis on the NINO3.4-controlled rainfall amount, frequency, and intensity fields (not shown).  Compared to the full-field EOFs, over much of peninsular India loadings are reduced in magnitude, especially for amount and frequency.  But otherwise, they are qualitatively similar to the corresponding non-NINO-controlled EOFs (Fig.~\ref{fig:eofs}): for amount the pattern is predominantly same-signed though with opposing behavior in the far eastern CMZ and just inland of the Western Ghats, for frequency the pattern is more homogeneous, and for intensity the WG-SEI-CMZ tripole is prominent.

Fig.~\ref{fig:nino34-regress-maps}(d-f) show the correlations between the NINO3.4-residual timeseries of DMI and JJAS rainfall amount, rainy day frequency, and mean rainfall intensity at each gridpoint.  For amount and frequency, the predominant signal is a WG-SEI-CMZ tripole, with positive IOD events moderately correlated with excess rain over most of the WG and CMZ and deficient rain over SEI.  Unlike for NINO3.4, the patterns in rainfall amount, rainy day frequency, and mean rainfall intensity are all grossly similar---to first order, all three are CMZ-SEI-WG tripoles, just with more positive values overall for rainy day frequency and more negative values overall for mean rainfall intensity.  Though the values over most gridpoints for all three are insignificant on statistical grounds, the correspondence of all three patterns to the physically well-grounded tripole mode suggests to us that they are unlikely artifacts.  Moreover, the tripole structure helps explain why the full, non-residual DMI timeseries correlation with AIRI is very small (\ilcorr{0.01}) but significantly larger for the tripole indicators (\ilcorr{0.31} for amount PC2, \ilcorr{0.34} for intensity PC1).

% \begin{figure*}[t]
%   \centering\noindent

%   \caption{Correlation coefficient in JJAS between the Dipole Mode Index (DMI) of the Indian Ocean Dipole with (a) rainfall, (b) rainy day frequency, and (c) mean rainfall intensity. Panels (d), (e), and (f) repeat (a), (b), and (c) but for each the fields are residuals after removing the NINO3.4 signal through linear regression.  \sah{Probably too many panels.  Maybe only show the NINO3.4 residuals (i.e. the bottom row)?}}
%   \label{fig:iod-corr-maps}
% \end{figure*}

% The NINO3.4-residual timeseries of DMI and AIRI are correlated at \ilcorr{0.20}, considerably higher than their full fields.
% amount PC2 (\ilcorr{0.31} full, \ilcorr{0.30} NINO3.4 residuals) and intensity PC1 (\ilcorr{0.34} full, \ilcorr{0.27} NINO3.4 residuals).
% Fig.~\ref{fig:nino34-regress-maps} shows correlation maps for the NINO3.4-controlled DMI with rainfall amount, frequency, and intensity.  Compared to the full fields, to first order for the NINO3.4 residuals the amount and frequency correlation patterns are unchanged in structure and are shifted quasi-homogeneously toward more positive values---with negative points becoming less negative or positive and positive points more positive---while the intensity correlations are weakly modified.  These differences from the full fields are broadly intuitive given the NINO3.4 correlation patterns (Fig.~\ref{fig:nino34-regress-maps}).

We conclude that, while it is the case that DMI is weakly correlated with AIRI, the IOD imprints a familiar spatial pattern, the WG-SEI-CMZ tripole.  Moreover, when the quasi-uniform drying influence of ENSO is linearly removed, the DMI correlations become quasi-uniformly more positive (even where they were negative), and as such the component of AIRI interannual variance not associated with ENSO is signficantly related to the IOD.

\section{Conclusions}
\label{sec:conc}

We use observational datasets of SST and high-resolution gridded rainfall spanning 1901-2020 to revisit three controversies in the literature involving relationships among the June-July-August-September (JJAS) seasonal averages of the All-India Rainfall Index (AIRI), the sub-India spatial distribution of rainfall variability, ENSO, and the Indian Ocean Dipole.
The first is the extent to which AIRI is representative of the spatiotemporal extent of rainfall anomalies.  Despite past concerns raised to the contrary, AIRI is found to be an excellent indicator of the spatial extent of wet anomalies across monsoonal India, and additionally of the average anomalous number of rainy days.  It is less closely related to the severity of rainfall anomalies when and where they occur.
% The EOF computed from standardized rainfall anomalies reflects this pattern, which also closely resembles the pointwise correlations with AIR, and its associated principal component timeseries also closely tracks that of AIR.
% Thus, despite rainfall variance being concentrated in the CMZ and WG, there is evidently a quasi-homogeneous signal on the seasonal timescale that acts on both these high-variance regions and most of the rest of monsoonal India.
% Years in which CMZ and WG are very wet (or dry) also tend to be years where the rest of India is, relative to local deviations, very wet (or dry).

The second is how AIRI variability relates to rainfall in fixed sub-India regions, and the third is how variability in AIRI and the subregions relate to ENSO and the IOD.  We confirm prior analyses of the leading variability mode via EOF analysis: it is a quasi-homogeneous wetting or drying that results primarily from an even more homogeneous and stronger signal in rainy day frequency, and it is strongly influenced by ENSO.  The second mode is a tripole with same-signed loadings in the high-rainfall Central Monsoon Zone and Western Ghats regions and opposite loadings in Southeastern India between.  This mode is associated more with rainfall intensity than with rainy day frequency and is effectively unrelated to ENSO.  The IOD imprints appreciably on the tripole mode and, with respect to the residual fields with the ENSO signal removed, imprints significantly on AIRI as well.

% \sah{Return to Mishra et al 2012 argument about preceding DJF ENSO influence on the 2nd mode.  Do we get that too?  Does DJF NINO predict JJAS DMI well, which would be one way that the two results could reconcile.}

Why is it that, across India, ENSO affects the number of rainy days strongly and fairly homogeneously?  This could reflect the increased static stability throughout the tropics driven by upper-tropospheric warming due to \elnino/ \cite{chiang_tropical_2002,su_teleconnection_2002}: convection becomes harder to initiate, resulting all else equal in fewer rainy days.  Why is the mean rainfall intensity increased over much of the CMZ with \elnino/ by contrast?  With weaker rainfall suppressed, more instability can build up in regions where it already is predilicted to, so that when convection comes it is intensified relative to neutral conditions.  This is analogous to the behavior of convection in coarse-resolution numerical models in which the convective parameterization is removed: large, very intense gridpoint storms emerge but much less frequently \cite{frierson_dynamics_2007,held_dynamic_2007}.

One controversy we have elided is the extent to which intraseasonal rainfall variability---from which in principle the seasonal-mean signal is removed---may nonlinearly rectify onto the seasonal-mean and thereby influence interannual variability \cite{goswami_intraseasonal_2001,lawrence_interannual_2001,krishnamurthy_seasonal_2008,qi_impacts_2009,sinha_leading_2011,moron_impact_2012}.  % and even centennial \cite{sinha_leading_2011} timescales.
This is largely orthogonal to the questions relating to lag-zero, JJAS-mean relationships of our present focus.  Nevertheless, the concern raised by this literature that the intraseasonal component is inherently chaotic and thus unpredictable \cite{palmer_chaos_1994,webster_monsoons:_1998} is worth keeping in mind, since the lag-zero correlations we have presented are partly means toward the end of improving the skill of seasonal forecasts of the summer monsoon rains.  At the same time, the well-documented relationships at the seasonal-mean timescale between the summer monsoon and ENSO, for example, convince us that clarifying the relationships between the monsoon and remote teleconnection modes is worthwhile, even if ultimately some portion of summer monsoon rainfall variance proves effectively unpredictable.

\acknowledgments
All authors acknowledge support from the Monsoon Mission Project under India’s Ministry of Earth Sciences.  S.A.H. acknowledges funding from the Columbia University Earth Institute Postdoctoral Fellowship.  We thank Sulochana Gadgil for many valuable discussions and comments.  We thank P.\@ Mukhopadhyay for helpful comments.
IMD rainfall data used is available at \url{https://www.imdpune.gov.in/Clim_Pred_LRF_New/Grided_Data_Download.html}.
% ERA5 data is available at \url{https://cds.climate.copernicus.eu/#!/search?text=ERA5&type=dataset}.
% NOAA OI SST data is available at \url{https://psl.noaa.gov/data/gridded/data.noaa.oisst.v2.html}.
ERSST data is available at \url{https://www.ncei.noaa.gov/pub/data/cmb/ersst/v5/netcdf/}

\bibliography{./references}

\begin{thebibliography}{}

\bibitem [\protect \citeauthoryear {%
Ashok%
, Feba%
\BCBL {}\ \BBA {} Tejavath%
}{%
Ashok%
\ \protect \BOthers {.}}{%
{\protect \APACyear {2019}}%
}]{%
ashok_indian_2019}
\APACinsertmetastar {%
ashok_indian_2019}%
\begin{APACrefauthors}%
Ashok, K.%
, Feba, F.%
\BCBL {}\ \BBA {} Tejavath, C\BPBI T.%
\end{APACrefauthors}%
\unskip\
\newblock
\APACrefYearMonthDay{2019}{{\APACmonth{07}}}{}.
\newblock
{\BBOQ}\APACrefatitle {The {{Indian}} Summer Monsoon Rainfall and {{ENSO}}}
  {The {{Indian}} summer monsoon rainfall and {{ENSO}}}.{\BBCQ}
\newblock
\APACjournalVolNumPages{Mausam}{70}{10}{443--452}.
\PrintBackRefs{\CurrentBib}

\bibitem [\protect \citeauthoryear {%
Ashok%
, Guan%
, Saji%
\BCBL {}\ \BBA {} Yamagata%
}{%
Ashok%
\ \protect \BOthers {.}}{%
{\protect \APACyear {2004}}%
}]{%
ashok_individual_2004}
\APACinsertmetastar {%
ashok_individual_2004}%
\begin{APACrefauthors}%
Ashok, K.%
, Guan, Z.%
, Saji, N\BPBI H.%
\BCBL {}\ \BBA {} Yamagata, T.%
\end{APACrefauthors}%
\unskip\
\newblock
\APACrefYearMonthDay{2004}{{\APACmonth{08}}}{}.
\newblock
{\BBOQ}\APACrefatitle {Individual and {{Combined Influences}} of {{ENSO}} and
  the {{Indian Ocean Dipole}} on the {{Indian Summer Monsoon}}} {Individual and
  {{Combined Influences}} of {{ENSO}} and the {{Indian Ocean Dipole}} on the
  {{Indian Summer Monsoon}}}.{\BBCQ}
\newblock
\APACjournalVolNumPages{J. Climate}{17}{16}{3141--3155}.
\newblock
\begin{APACrefDOI} \doi{10.1175/1520-0442(2004)017<3141:IACIOE>2.0.CO;2}
  \end{APACrefDOI}
\PrintBackRefs{\CurrentBib}

\bibitem [\protect \citeauthoryear {%
Ashok%
, Guan%
\BCBL {}\ \BBA {} Yamagata%
}{%
Ashok%
\ \protect \BOthers {.}}{%
{\protect \APACyear {2001}}%
}]{%
ashok_impact_2001}
\APACinsertmetastar {%
ashok_impact_2001}%
\begin{APACrefauthors}%
Ashok, K.%
, Guan, Z.%
\BCBL {}\ \BBA {} Yamagata, T.%
\end{APACrefauthors}%
\unskip\
\newblock
\APACrefYearMonthDay{2001}{}{}.
\newblock
{\BBOQ}\APACrefatitle {Impact of the {{Indian Ocean}} Dipole on the
  Relationship between the {{Indian}} Monsoon Rainfall and {{ENSO}}} {Impact of
  the {{Indian Ocean}} dipole on the relationship between the {{Indian}}
  monsoon rainfall and {{ENSO}}}.{\BBCQ}
\newblock
\APACjournalVolNumPages{Geophysical Research Letters}{28}{23}{4499--4502}.
\newblock
\begin{APACrefDOI} \doi{10.1029/2001GL013294} \end{APACrefDOI}
\PrintBackRefs{\CurrentBib}

\bibitem [\protect \citeauthoryear {%
Ashok%
\ \BBA {} Saji%
}{%
Ashok%
\ \BBA {} Saji%
}{%
{\protect \APACyear {2007}}%
}]{%
ashok_impacts_2007}
\APACinsertmetastar {%
ashok_impacts_2007}%
\begin{APACrefauthors}%
Ashok, K.%
\BCBT {}\ \BBA {} Saji, N\BPBI H.%
\end{APACrefauthors}%
\unskip\
\newblock
\APACrefYearMonthDay{2007}{{\APACmonth{08}}}{}.
\newblock
{\BBOQ}\APACrefatitle {On the Impacts of {{ENSO}} and {{Indian Ocean}} Dipole
  Events on Sub-Regional {{Indian}} Summer Monsoon Rainfall} {On the impacts of
  {{ENSO}} and {{Indian Ocean}} dipole events on sub-regional {{Indian}} summer
  monsoon rainfall}.{\BBCQ}
\newblock
\APACjournalVolNumPages{Nat Hazards}{42}{2}{273--285}.
\newblock
\begin{APACrefDOI} \doi{10.1007/s11069-006-9091-0} \end{APACrefDOI}
\PrintBackRefs{\CurrentBib}

\bibitem [\protect \citeauthoryear {%
Chiang%
\ \BBA {} Sobel%
}{%
Chiang%
\ \BBA {} Sobel%
}{%
{\protect \APACyear {2002}}%
}]{%
chiang_tropical_2002}
\APACinsertmetastar {%
chiang_tropical_2002}%
\begin{APACrefauthors}%
Chiang, J\BPBI C\BPBI H.%
\BCBT {}\ \BBA {} Sobel, A\BPBI H.%
\end{APACrefauthors}%
\unskip\
\newblock
\APACrefYearMonthDay{2002}{{\APACmonth{09}}}{}.
\newblock
{\BBOQ}\APACrefatitle {Tropical {{Tropospheric Temperature Variations Caused}}
  by {{ENSO}} and {{Their Influence}} on the {{Remote Tropical Climate}}}
  {Tropical {{Tropospheric Temperature Variations Caused}} by {{ENSO}} and
  {{Their Influence}} on the {{Remote Tropical Climate}}}.{\BBCQ}
\newblock
\APACjournalVolNumPages{J. Climate}{15}{18}{2616--2631}.
\newblock
\begin{APACrefDOI} \doi{10.1175/1520-0442(2002)015<2616:TTTVCB>2.0.CO;2}
  \end{APACrefDOI}
\PrintBackRefs{\CurrentBib}

\bibitem [\protect \citeauthoryear {%
Flynn%
, Nesbitt%
, Anders%
\BCBL {}\ \BBA {} Garg%
}{%
Flynn%
\ \protect \BOthers {.}}{%
{\protect \APACyear {2017}}%
}]{%
flynn_mesoscale_2017}
\APACinsertmetastar {%
flynn_mesoscale_2017}%
\begin{APACrefauthors}%
Flynn, W\BPBI J.%
, Nesbitt, S\BPBI W.%
, Anders, A\BPBI M.%
\BCBL {}\ \BBA {} Garg, P.%
\end{APACrefauthors}%
\unskip\
\newblock
\APACrefYearMonthDay{2017}{}{}.
\newblock
{\BBOQ}\APACrefatitle {Mesoscale Precipitation Characteristics near the
  {{Western Ghats}} during the {{Indian Summer Monsoon}} as Simulated by a
  High-Resolution Regional Model} {Mesoscale precipitation characteristics near
  the {{Western Ghats}} during the {{Indian Summer Monsoon}} as simulated by a
  high-resolution regional model}.{\BBCQ}
\newblock
\APACjournalVolNumPages{Quarterly Journal of the Royal Meteorological
  Society}{143}{709}{3070--3084}.
\newblock
\begin{APACrefDOI} \doi{10.1002/qj.3163} \end{APACrefDOI}
\PrintBackRefs{\CurrentBib}

\bibitem [\protect \citeauthoryear {%
Francis%
\ \BBA {} Gadgil%
}{%
Francis%
\ \BBA {} Gadgil%
}{%
{\protect \APACyear {2006}}%
}]{%
francis_intense_2006}
\APACinsertmetastar {%
francis_intense_2006}%
\begin{APACrefauthors}%
Francis, P\BPBI A.%
\BCBT {}\ \BBA {} Gadgil, S.%
\end{APACrefauthors}%
\unskip\
\newblock
\APACrefYearMonthDay{2006}{{\APACmonth{11}}}{}.
\newblock
{\BBOQ}\APACrefatitle {Intense Rainfall Events over the West Coast of
  {{India}}} {Intense rainfall events over the west coast of {{India}}}.{\BBCQ}
\newblock
\APACjournalVolNumPages{Meteorol. Atmos. Phys.}{94}{1}{27--42}.
\newblock
\begin{APACrefDOI} \doi{10.1007/s00703-005-0167-2} \end{APACrefDOI}
\PrintBackRefs{\CurrentBib}

\bibitem [\protect \citeauthoryear {%
Frierson%
}{%
Frierson%
}{%
{\protect \APACyear {2007}}%
}]{%
frierson_dynamics_2007}
\APACinsertmetastar {%
frierson_dynamics_2007}%
\begin{APACrefauthors}%
Frierson, D\BPBI M\BPBI W.%
\end{APACrefauthors}%
\unskip\
\newblock
\APACrefYearMonthDay{2007}{{\APACmonth{06}}}{}.
\newblock
{\BBOQ}\APACrefatitle {The Dynamics of Idealized Convection Schemes and Their
  Effect on the Zonally Averaged Tropical Circulation} {The dynamics of
  idealized convection schemes and their effect on the zonally averaged
  tropical circulation}.{\BBCQ}
\newblock
\APACjournalVolNumPages{J. Atmos. Sci.}{64}{6}{1959--1976}.
\newblock
\begin{APACrefDOI} \doi{10.1175/JAS3935.1} \end{APACrefDOI}
\PrintBackRefs{\CurrentBib}

\bibitem [\protect \citeauthoryear {%
Gadgil%
}{%
Gadgil%
}{%
{\protect \APACyear {1995}}%
}]{%
gadgil_climate_1995}
\APACinsertmetastar {%
gadgil_climate_1995}%
\begin{APACrefauthors}%
Gadgil, S.%
\end{APACrefauthors}%
\unskip\
\newblock
\APACrefYearMonthDay{1995}{}{}.
\newblock
{\BBOQ}\APACrefatitle {Climate Change and Agriculture \textendash{} {{An
  Indian}} Perspective} {Climate change and agriculture \textendash{} {{An
  Indian}} perspective}.{\BBCQ}
\newblock
\APACjournalVolNumPages{Current Science}{69}{8}{649--659}.
\PrintBackRefs{\CurrentBib}

\bibitem [\protect \citeauthoryear {%
Gadgil%
\ \BBA {} Gadgil%
}{%
Gadgil%
\ \BBA {} Gadgil%
}{%
{\protect \APACyear {2006}}%
}]{%
gadgil_indian_2006}
\APACinsertmetastar {%
gadgil_indian_2006}%
\begin{APACrefauthors}%
Gadgil, S.%
\BCBT {}\ \BBA {} Gadgil, S.%
\end{APACrefauthors}%
\unskip\
\newblock
\APACrefYearMonthDay{2006}{}{}.
\newblock
{\BBOQ}\APACrefatitle {The {{Indian Monsoon}}, {{GDP}} and {{Agriculture}}}
  {The {{Indian Monsoon}}, {{GDP}} and {{Agriculture}}}.{\BBCQ}
\newblock
\APACjournalVolNumPages{Economic and Political Weekly}{41}{47}{4887--4895}.
\PrintBackRefs{\CurrentBib}

\bibitem [\protect \citeauthoryear {%
Gadgil%
, Rajendran%
\BCBL {}\ \BBA {} Pai%
}{%
Gadgil%
\ \protect \BOthers {.}}{%
{\protect \APACyear {2019}}%
}]{%
gadgil_new_2019}
\APACinsertmetastar {%
gadgil_new_2019}%
\begin{APACrefauthors}%
Gadgil, S.%
, Rajendran, K.%
\BCBL {}\ \BBA {} Pai, D\BPBI S.%
\end{APACrefauthors}%
\unskip\
\newblock
\APACrefYearMonthDay{2019}{{\APACmonth{07}}}{}.
\newblock
{\BBOQ}\APACrefatitle {A New Rain-Based Index for the {{Indian}} Summer Monsoon
  Rainfall} {A new rain-based index for the {{Indian}} summer monsoon
  rainfall}.{\BBCQ}
\newblock
\APACjournalVolNumPages{Mausam}{70}{3}{485--500}.
\PrintBackRefs{\CurrentBib}

\bibitem [\protect \citeauthoryear {%
Gadgil%
, Vinayachandran%
, Francis%
\BCBL {}\ \BBA {} Gadgil%
}{%
Gadgil%
\ \protect \BOthers {.}}{%
{\protect \APACyear {2004}}%
}]{%
gadgil_extremes_2004}
\APACinsertmetastar {%
gadgil_extremes_2004}%
\begin{APACrefauthors}%
Gadgil, S.%
, Vinayachandran, P\BPBI N.%
, Francis, P\BPBI A.%
\BCBL {}\ \BBA {} Gadgil, S.%
\end{APACrefauthors}%
\unskip\
\newblock
\APACrefYearMonthDay{2004}{{\APACmonth{06}}}{}.
\newblock
{\BBOQ}\APACrefatitle {Extremes of the {{Indian}} Summer Monsoon Rainfall,
  {{ENSO}} and Equatorial {{Indian Ocean}} Oscillation} {Extremes of the
  {{Indian}} summer monsoon rainfall, {{ENSO}} and equatorial {{Indian Ocean}}
  oscillation}.{\BBCQ}
\newblock
\APACjournalVolNumPages{Geophysical Research Letters}{31}{12}{L12213}.
\newblock
\begin{APACrefDOI} \doi{10.1029/2004GL019733} \end{APACrefDOI}
\PrintBackRefs{\CurrentBib}

\bibitem [\protect \citeauthoryear {%
Gadgil%
, Yadumani%
\BCBL {}\ \BBA {} Joshi%
}{%
Gadgil%
\ \protect \BOthers {.}}{%
{\protect \APACyear {1993}}%
}]{%
gadgil_coherent_1993}
\APACinsertmetastar {%
gadgil_coherent_1993}%
\begin{APACrefauthors}%
Gadgil, S.%
, Yadumani%
\BCBL {}\ \BBA {} Joshi, N\BPBI V.%
\end{APACrefauthors}%
\unskip\
\newblock
\APACrefYearMonthDay{1993}{}{}.
\newblock
{\BBOQ}\APACrefatitle {Coherent Rainfall Zones of the {{Indian}} Region}
  {Coherent rainfall zones of the {{Indian}} region}.{\BBCQ}
\newblock
\APACjournalVolNumPages{International Journal of Climatology}{13}{5}{547--566}.
\newblock
\begin{APACrefDOI} \doi{10.1002/joc.3370130506} \end{APACrefDOI}
\PrintBackRefs{\CurrentBib}

\bibitem [\protect \citeauthoryear {%
Goswami%
\ \BBA {} Ajaya~Mohan%
}{%
Goswami%
\ \BBA {} Ajaya~Mohan%
}{%
{\protect \APACyear {2001}}%
}]{%
goswami_intraseasonal_2001}
\APACinsertmetastar {%
goswami_intraseasonal_2001}%
\begin{APACrefauthors}%
Goswami, B\BPBI N.%
\BCBT {}\ \BBA {} Ajaya~Mohan, R\BPBI S.%
\end{APACrefauthors}%
\unskip\
\newblock
\APACrefYearMonthDay{2001}{{\APACmonth{03}}}{}.
\newblock
{\BBOQ}\APACrefatitle {Intraseasonal {{Oscillations}} and {{Interannual
  Variability}} of the {{Indian Summer Monsoon}}} {Intraseasonal
  {{Oscillations}} and {{Interannual Variability}} of the {{Indian Summer
  Monsoon}}}.{\BBCQ}
\newblock
\APACjournalVolNumPages{Journal of Climate}{14}{6}{1180--1198}.
\newblock
\begin{APACrefDOI} \doi{10.1175/1520-0442(2001)014<1180:IOAIVO>2.0.CO;2}
  \end{APACrefDOI}
\PrintBackRefs{\CurrentBib}

\bibitem [\protect \citeauthoryear {%
Held%
, Zhao%
\BCBL {}\ \BBA {} Wyman%
}{%
Held%
\ \protect \BOthers {.}}{%
{\protect \APACyear {2007}}%
}]{%
held_dynamic_2007}
\APACinsertmetastar {%
held_dynamic_2007}%
\begin{APACrefauthors}%
Held, I\BPBI M.%
, Zhao, M.%
\BCBL {}\ \BBA {} Wyman, B.%
\end{APACrefauthors}%
\unskip\
\newblock
\APACrefYearMonthDay{2007}{{\APACmonth{01}}}{}.
\newblock
{\BBOQ}\APACrefatitle {Dynamic {{Radiative}}\textendash{{Convective Equilibria
  Using GCM Column Physics}}} {Dynamic {{Radiative}}\textendash{{Convective
  Equilibria Using GCM Column Physics}}}.{\BBCQ}
\newblock
\APACjournalVolNumPages{J. Atmos. Sci.}{64}{1}{228--238}.
\newblock
\begin{APACrefDOI} \doi{10.1175/JAS3825.11} \end{APACrefDOI}
\PrintBackRefs{\CurrentBib}

\bibitem [\protect \citeauthoryear {%
Huang%
\ \protect \BOthers {.}}{%
Huang%
\ \protect \BOthers {.}}{%
{\protect \APACyear {2015}}%
}]{%
huang_extended_2015}
\APACinsertmetastar {%
huang_extended_2015}%
\begin{APACrefauthors}%
Huang, B.%
, Banzon, V\BPBI F.%
, Freeman, E.%
, Lawrimore, J.%
, Liu, W.%
, Peterson, T\BPBI C.%
\BDBL {}Zhang, H\BHBI M.%
\end{APACrefauthors}%
\unskip\
\newblock
\APACrefYearMonthDay{2015}{{\APACmonth{02}}}{}.
\newblock
{\BBOQ}\APACrefatitle {Extended {{Reconstructed Sea Surface Temperature
  Version}} 4 ({{ERSST}}.v4). {{Part I}}: Upgrades and {{Intercomparisons}}}
  {Extended {{Reconstructed Sea Surface Temperature Version}} 4 ({{ERSST}}.v4).
  {{Part I}}: Upgrades and {{Intercomparisons}}}.{\BBCQ}
\newblock
\APACjournalVolNumPages{J. Climate}{28}{3}{911--930}.
\newblock
\begin{APACrefDOI} \doi{10.1175/JCLI-D-14-00006.1} \end{APACrefDOI}
\PrintBackRefs{\CurrentBib}

\bibitem [\protect \citeauthoryear {%
Huffman%
\ \protect \BOthers {.}}{%
Huffman%
\ \protect \BOthers {.}}{%
{\protect \APACyear {2007}}%
}]{%
huffman_trmm_2007}
\APACinsertmetastar {%
huffman_trmm_2007}%
\begin{APACrefauthors}%
Huffman, G\BPBI J.%
, Bolvin, D\BPBI T.%
, Nelkin, E\BPBI J.%
, Wolff, D\BPBI B.%
, Adler, R\BPBI F.%
, Gu, G.%
\BDBL {}Stocker, E\BPBI F.%
\end{APACrefauthors}%
\unskip\
\newblock
\APACrefYearMonthDay{2007}{{\APACmonth{02}}}{}.
\newblock
{\BBOQ}\APACrefatitle {The {{TRMM Multisatellite Precipitation Analysis}}
  ({{TMPA}}): Quasi-{{Global}}, {{Multiyear}}, {{Combined}}-{{Sensor
  Precipitation Estimates}} at {{Fine Scales}}} {The {{TRMM Multisatellite
  Precipitation Analysis}} ({{TMPA}}): Quasi-{{Global}}, {{Multiyear}},
  {{Combined}}-{{Sensor Precipitation Estimates}} at {{Fine Scales}}}.{\BBCQ}
\newblock
\APACjournalVolNumPages{J. Hydrometeor}{8}{1}{38--55}.
\newblock
\begin{APACrefDOI} \doi{10.1175/JHM560.1} \end{APACrefDOI}
\PrintBackRefs{\CurrentBib}

\bibitem [\protect \citeauthoryear {%
Hunt%
, Turner%
, Stein%
, Fletcher%
\BCBL {}\ \BBA {} Schiemann%
}{%
Hunt%
\ \protect \BOthers {.}}{%
{\protect \APACyear {2021}}%
}]{%
hunt_modes_2021}
\APACinsertmetastar {%
hunt_modes_2021}%
\begin{APACrefauthors}%
Hunt, K\BPBI M\BPBI R.%
, Turner, A\BPBI G.%
, Stein, T\BPBI H\BPBI M.%
, Fletcher, J\BPBI K.%
\BCBL {}\ \BBA {} Schiemann, R\BPBI K\BPBI H.%
\end{APACrefauthors}%
\unskip\
\newblock
\APACrefYearMonthDay{2021}{}{}.
\newblock
{\BBOQ}\APACrefatitle {Modes of Coastal Precipitation over Southwest {{India}}
  and Their Relationship with Intraseasonal Variability} {Modes of coastal
  precipitation over southwest {{India}} and their relationship with
  intraseasonal variability}.{\BBCQ}
\newblock
\APACjournalVolNumPages{Quarterly Journal of the Royal Meteorological
  Society}{147}{734}{181--201}.
\newblock
\begin{APACrefDOI} \doi{10.1002/qj.3913} \end{APACrefDOI}
\PrintBackRefs{\CurrentBib}

\bibitem [\protect \citeauthoryear {%
Ihara%
, Kushnir%
, Cane%
\BCBL {}\ \BBA {} Pe{\~n}a%
}{%
Ihara%
\ \protect \BOthers {.}}{%
{\protect \APACyear {2007}}%
}]{%
ihara_indian_2007}
\APACinsertmetastar {%
ihara_indian_2007}%
\begin{APACrefauthors}%
Ihara, C.%
, Kushnir, Y.%
, Cane, M\BPBI A.%
\BCBL {}\ \BBA {} Pe{\~n}a, V\BPBI H\BPBI D\BPBI L.%
\end{APACrefauthors}%
\unskip\
\newblock
\APACrefYearMonthDay{2007}{}{}.
\newblock
{\BBOQ}\APACrefatitle {Indian Summer Monsoon Rainfall and Its Link with
  {{ENSO}} and {{Indian Ocean}} Climate Indices} {Indian summer monsoon
  rainfall and its link with {{ENSO}} and {{Indian Ocean}} climate
  indices}.{\BBCQ}
\newblock
\APACjournalVolNumPages{International Journal of Climatology}{27}{2}{179--187}.
\newblock
\begin{APACrefDOI} \doi{10.1002/joc.1394} \end{APACrefDOI}
\PrintBackRefs{\CurrentBib}

\bibitem [\protect \citeauthoryear {%
Kelkar%
\ \BBA {} Sreejith%
}{%
Kelkar%
\ \BBA {} Sreejith%
}{%
{\protect \APACyear {2020}}%
}]{%
kelkar_meteorological_2020}
\APACinsertmetastar {%
kelkar_meteorological_2020}%
\begin{APACrefauthors}%
Kelkar, R\BPBI R.%
\BCBT {}\ \BBA {} Sreejith, O\BPBI P.%
\end{APACrefauthors}%
\unskip\
\newblock
\APACrefYearMonthDay{2020}{{\APACmonth{10}}}{}.
\newblock
{\BBOQ}\APACrefatitle {Meteorological Sub-Divisions of {{India}} and Their
  Geopolitical Evolution from 1875 to 2020} {Meteorological sub-divisions of
  {{India}} and their geopolitical evolution from 1875 to 2020}.{\BBCQ}
\newblock
\APACjournalVolNumPages{Mausam}{71}{4}{571--584}.
\PrintBackRefs{\CurrentBib}

\bibitem [\protect \citeauthoryear {%
Krishnamurthy%
\ \BBA {} Kirtman%
}{%
Krishnamurthy%
\ \BBA {} Kirtman%
}{%
{\protect \APACyear {2003}}%
}]{%
krishnamurthy_variability_2003}
\APACinsertmetastar {%
krishnamurthy_variability_2003}%
\begin{APACrefauthors}%
Krishnamurthy, V.%
\BCBT {}\ \BBA {} Kirtman, B\BPBI P.%
\end{APACrefauthors}%
\unskip\
\newblock
\APACrefYearMonthDay{2003}{}{}.
\newblock
{\BBOQ}\APACrefatitle {Variability of the {{Indian Ocean}}: Relation to Monsoon
  and {{ENSO}}} {Variability of the {{Indian Ocean}}: Relation to monsoon and
  {{ENSO}}}.{\BBCQ}
\newblock
\APACjournalVolNumPages{Quarterly Journal of the Royal Meteorological
  Society}{129}{590}{1623--1646}.
\newblock
\begin{APACrefDOI} \doi{10.1256/qj.01.166} \end{APACrefDOI}
\PrintBackRefs{\CurrentBib}

\bibitem [\protect \citeauthoryear {%
Krishnamurthy%
\ \BBA {} Shukla%
}{%
Krishnamurthy%
\ \BBA {} Shukla%
}{%
{\protect \APACyear {2000}}%
}]{%
krishnamurthy_intraseasonal_2000}
\APACinsertmetastar {%
krishnamurthy_intraseasonal_2000}%
\begin{APACrefauthors}%
Krishnamurthy, V.%
\BCBT {}\ \BBA {} Shukla, J.%
\end{APACrefauthors}%
\unskip\
\newblock
\APACrefYearMonthDay{2000}{{\APACmonth{12}}}{}.
\newblock
{\BBOQ}\APACrefatitle {Intraseasonal and {{Interannual Variability}} of
  {{Rainfall}} over {{India}}} {Intraseasonal and {{Interannual Variability}}
  of {{Rainfall}} over {{India}}}.{\BBCQ}
\newblock
\APACjournalVolNumPages{J. Climate}{13}{24}{4366--4377}.
\newblock
\begin{APACrefDOI} \doi{10.1175/1520-0442(2000)013<0001:IAIVOR>2.0.CO;2}
  \end{APACrefDOI}
\PrintBackRefs{\CurrentBib}

\bibitem [\protect \citeauthoryear {%
Krishnamurthy%
\ \BBA {} Shukla%
}{%
Krishnamurthy%
\ \BBA {} Shukla%
}{%
{\protect \APACyear {2007}}%
}]{%
krishnamurthy_intraseasonal_2007}
\APACinsertmetastar {%
krishnamurthy_intraseasonal_2007}%
\begin{APACrefauthors}%
Krishnamurthy, V.%
\BCBT {}\ \BBA {} Shukla, J.%
\end{APACrefauthors}%
\unskip\
\newblock
\APACrefYearMonthDay{2007}{{\APACmonth{01}}}{}.
\newblock
{\BBOQ}\APACrefatitle {Intraseasonal and {{Seasonally Persisting Patterns}} of
  {{Indian Monsoon Rainfall}}} {Intraseasonal and {{Seasonally Persisting
  Patterns}} of {{Indian Monsoon Rainfall}}}.{\BBCQ}
\newblock
\APACjournalVolNumPages{J. Climate}{20}{1}{3--20}.
\newblock
\begin{APACrefDOI} \doi{10.1175/JCLI3981.1} \end{APACrefDOI}
\PrintBackRefs{\CurrentBib}

\bibitem [\protect \citeauthoryear {%
Krishnamurthy%
\ \BBA {} Shukla%
}{%
Krishnamurthy%
\ \BBA {} Shukla%
}{%
{\protect \APACyear {2008}}%
}]{%
krishnamurthy_seasonal_2008}
\APACinsertmetastar {%
krishnamurthy_seasonal_2008}%
\begin{APACrefauthors}%
Krishnamurthy, V.%
\BCBT {}\ \BBA {} Shukla, J.%
\end{APACrefauthors}%
\unskip\
\newblock
\APACrefYearMonthDay{2008}{{\APACmonth{03}}}{}.
\newblock
{\BBOQ}\APACrefatitle {Seasonal Persistence and Propagation of Intraseasonal
  Patterns over the {{Indian}} Monsoon Region} {Seasonal persistence and
  propagation of intraseasonal patterns over the {{Indian}} monsoon
  region}.{\BBCQ}
\newblock
\APACjournalVolNumPages{Clim Dyn}{30}{4}{353--369}.
\newblock
\begin{APACrefDOI} \doi{10.1007/s00382-007-0300-7} \end{APACrefDOI}
\PrintBackRefs{\CurrentBib}

\bibitem [\protect \citeauthoryear {%
Kumar%
, Rajagopalan%
\BCBL {}\ \BBA {} Cane%
}{%
Kumar%
\ \protect \BOthers {.}}{%
{\protect \APACyear {1999}}%
}]{%
kumar_weakening_1999}
\APACinsertmetastar {%
kumar_weakening_1999}%
\begin{APACrefauthors}%
Kumar, K\BPBI K.%
, Rajagopalan, B.%
\BCBL {}\ \BBA {} Cane, M\BPBI A.%
\end{APACrefauthors}%
\unskip\
\newblock
\APACrefYearMonthDay{1999}{{\APACmonth{06}}}{}.
\newblock
{\BBOQ}\APACrefatitle {On the {{Weakening Relationship Between}} the {{Indian
  Monsoon}} and {{ENSO}}} {On the {{Weakening Relationship Between}} the
  {{Indian Monsoon}} and {{ENSO}}}.{\BBCQ}
\newblock
\APACjournalVolNumPages{Science}{284}{5423}{2156--2159}.
\newblock
\begin{APACrefDOI} \doi{10.1126/science.284.5423.2156} \end{APACrefDOI}
\PrintBackRefs{\CurrentBib}

\bibitem [\protect \citeauthoryear {%
Lawrence%
\ \BBA {} Webster%
}{%
Lawrence%
\ \BBA {} Webster%
}{%
{\protect \APACyear {2001}}%
}]{%
lawrence_interannual_2001}
\APACinsertmetastar {%
lawrence_interannual_2001}%
\begin{APACrefauthors}%
Lawrence, D\BPBI M.%
\BCBT {}\ \BBA {} Webster, P\BPBI J.%
\end{APACrefauthors}%
\unskip\
\newblock
\APACrefYearMonthDay{2001}{{\APACmonth{07}}}{}.
\newblock
{\BBOQ}\APACrefatitle {Interannual {{Variations}} of the {{Intraseasonal
  Oscillation}} in the {{South Asian Summer Monsoon Region}}} {Interannual
  {{Variations}} of the {{Intraseasonal Oscillation}} in the {{South Asian
  Summer Monsoon Region}}}.{\BBCQ}
\newblock
\APACjournalVolNumPages{Journal of Climate}{14}{13}{2910--2922}.
\newblock
\begin{APACrefDOI} \doi{10.1175/1520-0442(2001)014<2910:IVOTIO>2.0.CO;2}
  \end{APACrefDOI}
\PrintBackRefs{\CurrentBib}

\bibitem [\protect \citeauthoryear {%
Mishra%
, Smoliak%
, Lettenmaier%
\BCBL {}\ \BBA {} Wallace%
}{%
Mishra%
\ \protect \BOthers {.}}{%
{\protect \APACyear {2012}}%
}]{%
mishra_prominent_2012}
\APACinsertmetastar {%
mishra_prominent_2012}%
\begin{APACrefauthors}%
Mishra, V.%
, Smoliak, B\BPBI V.%
, Lettenmaier, D\BPBI P.%
\BCBL {}\ \BBA {} Wallace, J\BPBI M.%
\end{APACrefauthors}%
\unskip\
\newblock
\APACrefYearMonthDay{2012}{{\APACmonth{05}}}{}.
\newblock
{\BBOQ}\APACrefatitle {A Prominent Pattern of Year-to-Year Variability in
  {{Indian Summer Monsoon Rainfall}}} {A prominent pattern of year-to-year
  variability in {{Indian Summer Monsoon Rainfall}}}.{\BBCQ}
\newblock
\APACjournalVolNumPages{PNAS}{109}{19}{7213--7217}.
\newblock
\begin{APACrefDOI} \doi{10.1073/pnas.1119150109} \end{APACrefDOI}
\PrintBackRefs{\CurrentBib}

\bibitem [\protect \citeauthoryear {%
Mooley%
, Parthasarathy%
, Sontakke%
\BCBL {}\ \BBA {} Munot%
}{%
Mooley%
\ \protect \BOthers {.}}{%
{\protect \APACyear {1981}}%
}]{%
mooley_annual_1981}
\APACinsertmetastar {%
mooley_annual_1981}%
\begin{APACrefauthors}%
Mooley, D\BPBI A.%
, Parthasarathy, B.%
, Sontakke, N\BPBI A.%
\BCBL {}\ \BBA {} Munot, A\BPBI A.%
\end{APACrefauthors}%
\unskip\
\newblock
\APACrefYearMonthDay{1981}{}{}.
\newblock
{\BBOQ}\APACrefatitle {Annual Rain-Water over {{India}}, Its Variability and
  Impact on the Economy} {Annual rain-water over {{India}}, its variability and
  impact on the economy}.{\BBCQ}
\newblock
\APACjournalVolNumPages{Journal of Climatology}{1}{2}{167--186}.
\newblock
\begin{APACrefDOI} \doi{10.1002/joc.3370010206} \end{APACrefDOI}
\PrintBackRefs{\CurrentBib}

\bibitem [\protect \citeauthoryear {%
Moron%
, Robertson%
\BCBL {}\ \BBA {} Ghil%
}{%
Moron%
\ \protect \BOthers {.}}{%
{\protect \APACyear {2012}}%
}]{%
moron_impact_2012}
\APACinsertmetastar {%
moron_impact_2012}%
\begin{APACrefauthors}%
Moron, V.%
, Robertson, A\BPBI W.%
\BCBL {}\ \BBA {} Ghil, M.%
\end{APACrefauthors}%
\unskip\
\newblock
\APACrefYearMonthDay{2012}{{\APACmonth{06}}}{}.
\newblock
{\BBOQ}\APACrefatitle {Impact of the Modulated Annual Cycle and Intraseasonal
  Oscillation on Daily-to-Interannual Rainfall Variability across Monsoonal
  {{India}}} {Impact of the modulated annual cycle and intraseasonal
  oscillation on daily-to-interannual rainfall variability across monsoonal
  {{India}}}.{\BBCQ}
\newblock
\APACjournalVolNumPages{Clim Dyn}{38}{11}{2409--2435}.
\newblock
\begin{APACrefDOI} \doi{10.1007/s00382-011-1253-4} \end{APACrefDOI}
\PrintBackRefs{\CurrentBib}

\bibitem [\protect \citeauthoryear {%
Moron%
, Robertson%
\BCBL {}\ \BBA {} Pai%
}{%
Moron%
\ \protect \BOthers {.}}{%
{\protect \APACyear {2017}}%
}]{%
moron_spatial_2017}
\APACinsertmetastar {%
moron_spatial_2017}%
\begin{APACrefauthors}%
Moron, V.%
, Robertson, A\BPBI W.%
\BCBL {}\ \BBA {} Pai, D\BPBI S.%
\end{APACrefauthors}%
\unskip\
\newblock
\APACrefYearMonthDay{2017}{{\APACmonth{11}}}{}.
\newblock
{\BBOQ}\APACrefatitle {On the Spatial Coherence of Sub-Seasonal to Seasonal
  {{Indian}} Rainfall Anomalies} {On the spatial coherence of sub-seasonal to
  seasonal {{Indian}} rainfall anomalies}.{\BBCQ}
\newblock
\APACjournalVolNumPages{Clim Dyn}{49}{9}{3403--3423}.
\newblock
\begin{APACrefDOI} \doi{10.1007/s00382-017-3520-5} \end{APACrefDOI}
\PrintBackRefs{\CurrentBib}

\bibitem [\protect \citeauthoryear {%
Pai%
\ \protect \BOthers {.}}{%
Pai%
\ \protect \BOthers {.}}{%
{\protect \APACyear {2014}}%
}]{%
pai_development_2014}
\APACinsertmetastar {%
pai_development_2014}%
\begin{APACrefauthors}%
Pai, D\BPBI S.%
, Sridhar, L.%
, Rajeevan, M.%
, Sreejith, O\BPBI P.%
, Satbhai, N\BPBI S.%
\BCBL {}\ \BBA {} Mukhopadhyay, B.%
\end{APACrefauthors}%
\unskip\
\newblock
\APACrefYearMonthDay{2014}{}{}.
\newblock
{\BBOQ}\APACrefatitle {Development of a New High Spatial Resolution
  (0.25\textdegree{} \texttimes{} 0.25\textdegree ) Long Period (1901-2010)
  Daily Gridded Rainfall Data Set over {{India}} and Its Comparison with
  Existing Data Sets over the Region} {Development of a new high spatial
  resolution (0.25\textdegree{} \texttimes{} 0.25\textdegree ) long period
  (1901-2010) daily gridded rainfall data set over {{India}} and its comparison
  with existing data sets over the region}.{\BBCQ}
\newblock
\APACjournalVolNumPages{Mausam}{65}{1}{18}.
\PrintBackRefs{\CurrentBib}

\bibitem [\protect \citeauthoryear {%
Palmer%
}{%
Palmer%
}{%
{\protect \APACyear {1994}}%
}]{%
palmer_chaos_1994}
\APACinsertmetastar {%
palmer_chaos_1994}%
\begin{APACrefauthors}%
Palmer, T\BPBI N.%
\end{APACrefauthors}%
\unskip\
\newblock
\APACrefYearMonthDay{1994}{}{}.
\newblock
{\BBOQ}\APACrefatitle {Chaos and Predictability in Forecasting the Monsoons}
  {Chaos and predictability in forecasting the monsoons}.{\BBCQ}
\newblock
\APACjournalVolNumPages{Proc. Indian Natl. Sci. Acd.}{60}{}{57--66}.
\PrintBackRefs{\CurrentBib}

\bibitem [\protect \citeauthoryear {%
Pant%
\ \BBA {} Parthasarathy%
}{%
Pant%
\ \BBA {} Parthasarathy%
}{%
{\protect \APACyear {1981}}%
}]{%
pant_aspects_1981}
\APACinsertmetastar {%
pant_aspects_1981}%
\begin{APACrefauthors}%
Pant, G\BPBI B.%
\BCBT {}\ \BBA {} Parthasarathy, S\BPBI B.%
\end{APACrefauthors}%
\unskip\
\newblock
\APACrefYearMonthDay{1981}{{\APACmonth{09}}}{}.
\newblock
{\BBOQ}\APACrefatitle {Some Aspects of an Association between the Southern
  Oscillation and Indian Summer Monsoon} {Some aspects of an association
  between the southern oscillation and indian summer monsoon}.{\BBCQ}
\newblock
\APACjournalVolNumPages{Arch. Met. Geoph. Biocl., Ser. B}{29}{3}{245--252}.
\newblock
\begin{APACrefDOI} \doi{10.1007/BF02263246} \end{APACrefDOI}
\PrintBackRefs{\CurrentBib}

\bibitem [\protect \citeauthoryear {%
Parthasarathy%
, Kumar%
\BCBL {}\ \BBA {} Munot%
}{%
Parthasarathy%
\ \protect \BOthers {.}}{%
{\protect \APACyear {1993}}%
}]{%
parthasarathy_homogeneous_1993}
\APACinsertmetastar {%
parthasarathy_homogeneous_1993}%
\begin{APACrefauthors}%
Parthasarathy, B.%
, Kumar, K\BPBI R.%
\BCBL {}\ \BBA {} Munot, A\BPBI A.%
\end{APACrefauthors}%
\unskip\
\newblock
\APACrefYearMonthDay{1993}{{\APACmonth{03}}}{}.
\newblock
{\BBOQ}\APACrefatitle {Homogeneous {{Indian Monsoon}} Rainfall: Variability and
  Prediction} {Homogeneous {{Indian Monsoon}} rainfall: Variability and
  prediction}.{\BBCQ}
\newblock
\APACjournalVolNumPages{Proc. Indian Acad. Sci. (Earth Planet
  Sci.)}{102}{1}{121--155}.
\newblock
\begin{APACrefDOI} \doi{10.1007/BF02839187} \end{APACrefDOI}
\PrintBackRefs{\CurrentBib}

\bibitem [\protect \citeauthoryear {%
Parthasarathy%
, Munot%
\BCBL {}\ \BBA {} Kothawale%
}{%
Parthasarathy%
\ \protect \BOthers {.}}{%
{\protect \APACyear {1988}}%
}]{%
parthasarathy_regression_1988}
\APACinsertmetastar {%
parthasarathy_regression_1988}%
\begin{APACrefauthors}%
Parthasarathy, B.%
, Munot, A\BPBI A.%
\BCBL {}\ \BBA {} Kothawale, D\BPBI R.%
\end{APACrefauthors}%
\unskip\
\newblock
\APACrefYearMonthDay{1988}{{\APACmonth{03}}}{}.
\newblock
{\BBOQ}\APACrefatitle {Regression Model for Estimation of Indian Foodgrain
  Production from Summer Monsoon Rainfall} {Regression model for estimation of
  indian foodgrain production from summer monsoon rainfall}.{\BBCQ}
\newblock
\APACjournalVolNumPages{Agricultural and Forest Meteorology}{42}{2}{167--182}.
\newblock
\begin{APACrefDOI} \doi{10.1016/0168-1923(88)90075-5} \end{APACrefDOI}
\PrintBackRefs{\CurrentBib}

\bibitem [\protect \citeauthoryear {%
Parthasarathy%
, Sontakke%
, Monot%
\BCBL {}\ \BBA {} Kothawale%
}{%
Parthasarathy%
\ \protect \BOthers {.}}{%
{\protect \APACyear {1987}}%
}]{%
parthasarathy_droughtsfloods_1987}
\APACinsertmetastar {%
parthasarathy_droughtsfloods_1987}%
\begin{APACrefauthors}%
Parthasarathy, B.%
, Sontakke, N\BPBI A.%
, Monot, A\BPBI A.%
\BCBL {}\ \BBA {} Kothawale, D\BPBI R.%
\end{APACrefauthors}%
\unskip\
\newblock
\APACrefYearMonthDay{1987}{}{}.
\newblock
{\BBOQ}\APACrefatitle {Droughts/Floods in the Summer Monsoon Season over
  Different Meteorological Subdivisions of {{India}} for the Period
  1871\textendash 1984} {Droughts/floods in the summer monsoon season over
  different meteorological subdivisions of {{India}} for the period
  1871\textendash 1984}.{\BBCQ}
\newblock
\APACjournalVolNumPages{Journal of Climatology}{7}{1}{57--70}.
\newblock
\begin{APACrefDOI} \doi{10.1002/joc.3370070106} \end{APACrefDOI}
\PrintBackRefs{\CurrentBib}

\bibitem [\protect \citeauthoryear {%
Qi%
, Zhang%
, Li%
\BCBL {}\ \BBA {} Wen%
}{%
Qi%
\ \protect \BOthers {.}}{%
{\protect \APACyear {2009}}%
}]{%
qi_impacts_2009}
\APACinsertmetastar {%
qi_impacts_2009}%
\begin{APACrefauthors}%
Qi, Y.%
, Zhang, R.%
, Li, T.%
\BCBL {}\ \BBA {} Wen, M.%
\end{APACrefauthors}%
\unskip\
\newblock
\APACrefYearMonthDay{2009}{{\APACmonth{03}}}{}.
\newblock
{\BBOQ}\APACrefatitle {Impacts of Intraseasonal Oscillation on the Onset and
  Interannual Variation of the {{Indian}} Summer Monsoon} {Impacts of
  intraseasonal oscillation on the onset and interannual variation of the
  {{Indian}} summer monsoon}.{\BBCQ}
\newblock
\APACjournalVolNumPages{Chin. Sci. Bull.}{54}{5}{880--884}.
\newblock
\begin{APACrefDOI} \doi{10.1007/s11434-008-0441-z} \end{APACrefDOI}
\PrintBackRefs{\CurrentBib}

\bibitem [\protect \citeauthoryear {%
Rajeevan%
, Gadgil%
\BCBL {}\ \BBA {} Bhate%
}{%
Rajeevan%
\ \protect \BOthers {.}}{%
{\protect \APACyear {2010}}%
}]{%
rajeevan_active_2010}
\APACinsertmetastar {%
rajeevan_active_2010}%
\begin{APACrefauthors}%
Rajeevan, M.%
, Gadgil, S.%
\BCBL {}\ \BBA {} Bhate, J.%
\end{APACrefauthors}%
\unskip\
\newblock
\APACrefYearMonthDay{2010}{{\APACmonth{06}}}{}.
\newblock
{\BBOQ}\APACrefatitle {Active and Break Spells of the {{Indian}} Summer
  Monsoon} {Active and break spells of the {{Indian}} summer monsoon}.{\BBCQ}
\newblock
\APACjournalVolNumPages{J Earth Syst Sci}{119}{3}{229--247}.
\newblock
\begin{APACrefDOI} \doi{10.1007/s12040-010-0019-4} \end{APACrefDOI}
\PrintBackRefs{\CurrentBib}

\bibitem [\protect \citeauthoryear {%
Ramesh%
, Nicolas%
\BCBL {}\ \BBA {} Boos%
}{%
Ramesh%
\ \protect \BOthers {.}}{%
{\protect \APACyear {2021}}%
}]{%
ramesh_globally_2021}
\APACinsertmetastar {%
ramesh_globally_2021}%
\begin{APACrefauthors}%
Ramesh, N.%
, Nicolas, Q.%
\BCBL {}\ \BBA {} Boos, W\BPBI R.%
\end{APACrefauthors}%
\unskip\
\newblock
\APACrefYearMonthDay{2021}{{\APACmonth{07}}}{}.
\newblock
{\BBOQ}\APACrefatitle {The {{Globally Coherent Pattern}} of {{Autumn Monsoon
  Precipitation}}} {The {{Globally Coherent Pattern}} of {{Autumn Monsoon
  Precipitation}}}.{\BBCQ}
\newblock
\APACjournalVolNumPages{Journal of Climate}{34}{14}{5687--5705}.
\newblock
\begin{APACrefDOI} \doi{10.1175/JCLI-D-20-0740.1} \end{APACrefDOI}
\PrintBackRefs{\CurrentBib}

\bibitem [\protect \citeauthoryear {%
Rasmusson%
\ \BBA {} Carpenter%
}{%
Rasmusson%
\ \BBA {} Carpenter%
}{%
{\protect \APACyear {1983}}%
}]{%
rasmusson_relationship_1983}
\APACinsertmetastar {%
rasmusson_relationship_1983}%
\begin{APACrefauthors}%
Rasmusson, E\BPBI M.%
\BCBT {}\ \BBA {} Carpenter, T\BPBI H.%
\end{APACrefauthors}%
\unskip\
\newblock
\APACrefYearMonthDay{1983}{{\APACmonth{03}}}{}.
\newblock
{\BBOQ}\APACrefatitle {The {{Relationship Between Eastern Equatorial Pacific
  Sea Surface Temperatures}} and {{Rainfall}} over {{India}} and {{Sri Lanka}}}
  {The {{Relationship Between Eastern Equatorial Pacific Sea Surface
  Temperatures}} and {{Rainfall}} over {{India}} and {{Sri Lanka}}}.{\BBCQ}
\newblock
\APACjournalVolNumPages{Mon. Wea. Rev.}{111}{3}{517--528}.
\newblock
\begin{APACrefDOI} \doi{10.1175/1520-0493(1983)111<0517:TRBEEP>2.0.CO;2}
  \end{APACrefDOI}
\PrintBackRefs{\CurrentBib}

\bibitem [\protect \citeauthoryear {%
Robertson%
\ \protect \BOthers {.}}{%
Robertson%
\ \protect \BOthers {.}}{%
{\protect \APACyear {2019}}%
}]{%
robertson_multi-scale_2019}
\APACinsertmetastar {%
robertson_multi-scale_2019}%
\begin{APACrefauthors}%
Robertson, A\BPBI W.%
, Moron, V.%
, Vigaud, N.%
, Acharya, N.%
, Greene, A\BPBI M.%
\BCBL {}\ \BBA {} Pai, D\BPBI S.%
\end{APACrefauthors}%
\unskip\
\newblock
\APACrefYearMonthDay{2019}{{\APACmonth{04}}}{}.
\newblock
{\BBOQ}\APACrefatitle {Multi-Scale Variability and Predictability of {{Indian}}
  Summer Monsoon Rainfall} {Multi-scale variability and predictability of
  {{Indian}} summer monsoon rainfall}.{\BBCQ}
\newblock
\APACjournalVolNumPages{Mausam}{70}{2}{277--292}.
\PrintBackRefs{\CurrentBib}

\bibitem [\protect \citeauthoryear {%
Saji%
, Goswami%
, Vinayachandran%
\BCBL {}\ \BBA {} Yamagata%
}{%
Saji%
\ \protect \BOthers {.}}{%
{\protect \APACyear {1999}}%
}]{%
saji_dipole_1999}
\APACinsertmetastar {%
saji_dipole_1999}%
\begin{APACrefauthors}%
Saji, N\BPBI H.%
, Goswami, B\BPBI N.%
, Vinayachandran, P\BPBI N.%
\BCBL {}\ \BBA {} Yamagata, T.%
\end{APACrefauthors}%
\unskip\
\newblock
\APACrefYearMonthDay{1999}{{\APACmonth{09}}}{}.
\newblock
{\BBOQ}\APACrefatitle {A Dipole Mode in the Tropical {{Indian Ocean}}} {A
  dipole mode in the tropical {{Indian Ocean}}}.{\BBCQ}
\newblock
\APACjournalVolNumPages{Nature}{401}{6751}{360--363}.
\newblock
\begin{APACrefDOI} \doi{10.1038/43854} \end{APACrefDOI}
\PrintBackRefs{\CurrentBib}

\bibitem [\protect \citeauthoryear {%
Shukla%
}{%
Shukla%
}{%
{\protect \APACyear {1987}}%
}]{%
shukla_interannual_1987}
\APACinsertmetastar {%
shukla_interannual_1987}%
\begin{APACrefauthors}%
Shukla, J.%
\end{APACrefauthors}%
\unskip\
\newblock
\APACrefYearMonthDay{1987}{}{}.
\newblock
{\BBOQ}\APACrefatitle {Interannual {{Variability}} of {{Monsoons}}}
  {Interannual {{Variability}} of {{Monsoons}}}.{\BBCQ}
\newblock
\BIn{} \APACrefbtitle {Monsoons} {Monsoons}\ (\BPGS\ 399--464).
\PrintBackRefs{\CurrentBib}

\bibitem [\protect \citeauthoryear {%
Sinha%
\ \protect \BOthers {.}}{%
Sinha%
\ \protect \BOthers {.}}{%
{\protect \APACyear {2011}}%
}]{%
sinha_leading_2011}
\APACinsertmetastar {%
sinha_leading_2011}%
\begin{APACrefauthors}%
Sinha, A.%
, Berkelhammer, M.%
, Stott, L.%
, Mudelsee, M.%
, Cheng, H.%
\BCBL {}\ \BBA {} Biswas, J.%
\end{APACrefauthors}%
\unskip\
\newblock
\APACrefYearMonthDay{2011}{}{}.
\newblock
{\BBOQ}\APACrefatitle {The Leading Mode of {{Indian Summer Monsoon}}
  Precipitation Variability during the Last Millennium} {The leading mode of
  {{Indian Summer Monsoon}} precipitation variability during the last
  millennium}.{\BBCQ}
\newblock
\APACjournalVolNumPages{Geophysical Research Letters}{38}{15}{}.
\newblock
\begin{APACrefDOI} \doi{10.1029/2011GL047713} \end{APACrefDOI}
\PrintBackRefs{\CurrentBib}

\bibitem [\protect \citeauthoryear {%
Straus%
\ \BBA {} Krishnamurthy%
}{%
Straus%
\ \BBA {} Krishnamurthy%
}{%
{\protect \APACyear {2007}}%
}]{%
straus_preferred_2007}
\APACinsertmetastar {%
straus_preferred_2007}%
\begin{APACrefauthors}%
Straus, D\BPBI M.%
\BCBT {}\ \BBA {} Krishnamurthy, V.%
\end{APACrefauthors}%
\unskip\
\newblock
\APACrefYearMonthDay{2007}{}{}.
\newblock
{\BBOQ}\APACrefatitle {The {{Preferred Structure}} of the {{Interannual Indian
  Monsoon Variability}}} {The {{Preferred Structure}} of the {{Interannual
  Indian Monsoon Variability}}}.{\BBCQ}
\newblock
\BIn{} M.~Sharan\ \BBA {} S.~Raman\ (\BEDS), \APACrefbtitle {Atmospheric and
  {{Oceanic}}: Mesoscale {{Processes}}} {Atmospheric and {{Oceanic}}: Mesoscale
  {{Processes}}}\ (\BPGS\ 1717--1732).
\newblock
\APACaddressPublisher{{Basel}}{{Birkh\"auser}}.
\newblock
\begin{APACrefDOI} \doi{10.1007/978-3-7643-8493-7_15} \end{APACrefDOI}
\PrintBackRefs{\CurrentBib}

\bibitem [\protect \citeauthoryear {%
Stuecker%
\ \protect \BOthers {.}}{%
Stuecker%
\ \protect \BOthers {.}}{%
{\protect \APACyear {2017}}%
}]{%
stuecker_revisiting_2017}
\APACinsertmetastar {%
stuecker_revisiting_2017}%
\begin{APACrefauthors}%
Stuecker, M\BPBI F.%
, Timmermann, A.%
, Jin, F\BHBI F.%
, Chikamoto, Y.%
, Zhang, W.%
, Wittenberg, A\BPBI T.%
\BDBL {}Zhao, S.%
\end{APACrefauthors}%
\unskip\
\newblock
\APACrefYearMonthDay{2017}{}{}.
\newblock
{\BBOQ}\APACrefatitle {Revisiting {{ENSO}}/{{Indian Ocean Dipole}} Phase
  Relationships} {Revisiting {{ENSO}}/{{Indian Ocean Dipole}} phase
  relationships}.{\BBCQ}
\newblock
\APACjournalVolNumPages{Geophysical Research Letters}{44}{5}{2481--2492}.
\newblock
\begin{APACrefDOI} \doi{10.1002/2016GL072308} \end{APACrefDOI}
\PrintBackRefs{\CurrentBib}

\bibitem [\protect \citeauthoryear {%
Su%
\ \BBA {} Neelin%
}{%
Su%
\ \BBA {} Neelin%
}{%
{\protect \APACyear {2002}}%
}]{%
su_teleconnection_2002}
\APACinsertmetastar {%
su_teleconnection_2002}%
\begin{APACrefauthors}%
Su, H.%
\BCBT {}\ \BBA {} Neelin, J\BPBI D.%
\end{APACrefauthors}%
\unskip\
\newblock
\APACrefYearMonthDay{2002}{{\APACmonth{09}}}{}.
\newblock
{\BBOQ}\APACrefatitle {Teleconnection {{Mechanisms}} for {{Tropical Pacific
  Descent Anomalies}} during {{El Ni\~no}}} {Teleconnection {{Mechanisms}} for
  {{Tropical Pacific Descent Anomalies}} during {{El Ni\~no}}}.{\BBCQ}
\newblock
\APACjournalVolNumPages{J. Atmos. Sci.}{59}{18}{2694--2712}.
\newblock
\begin{APACrefDOI} \doi{10.1175/1520-0469(2002)059<2694:TMFTPD>2.0.CO;2}
  \end{APACrefDOI}
\PrintBackRefs{\CurrentBib}

\bibitem [\protect \citeauthoryear {%
Surendran%
, Gadgil%
, Francis%
\BCBL {}\ \BBA {} Rajeevan%
}{%
Surendran%
\ \protect \BOthers {.}}{%
{\protect \APACyear {2015}}%
}]{%
surendran_prediction_2015}
\APACinsertmetastar {%
surendran_prediction_2015}%
\begin{APACrefauthors}%
Surendran, S.%
, Gadgil, S.%
, Francis, P\BPBI A.%
\BCBL {}\ \BBA {} Rajeevan, M.%
\end{APACrefauthors}%
\unskip\
\newblock
\APACrefYearMonthDay{2015}{{\APACmonth{09}}}{}.
\newblock
{\BBOQ}\APACrefatitle {Prediction of {{Indian}} Rainfall during the Summer
  Monsoon Season on the Basis of Links with Equatorial {{Pacific}} and {{Indian
  Ocean}} Climate Indices} {Prediction of {{Indian}} rainfall during the summer
  monsoon season on the basis of links with equatorial {{Pacific}} and {{Indian
  Ocean}} climate indices}.{\BBCQ}
\newblock
\APACjournalVolNumPages{Environ. Res. Lett.}{10}{9}{094004}.
\newblock
\begin{APACrefDOI} \doi{10.1088/1748-9326/10/9/094004} \end{APACrefDOI}
\PrintBackRefs{\CurrentBib}

\bibitem [\protect \citeauthoryear {%
Vecchi%
\ \BBA {} Harrison%
}{%
Vecchi%
\ \BBA {} Harrison%
}{%
{\protect \APACyear {2004}}%
}]{%
vecchi_interannual_2004}
\APACinsertmetastar {%
vecchi_interannual_2004}%
\begin{APACrefauthors}%
Vecchi, G\BPBI A.%
\BCBT {}\ \BBA {} Harrison, D\BPBI E.%
\end{APACrefauthors}%
\unskip\
\newblock
\APACrefYearMonthDay{2004}{}{}.
\newblock
{\BBOQ}\APACrefatitle {Interannual {{Indian Rainfall Variability}} and {{Indian
  Ocean Sea Surface Temperature Anomalies}}} {Interannual {{Indian Rainfall
  Variability}} and {{Indian Ocean Sea Surface Temperature Anomalies}}}.{\BBCQ}
\newblock
\BIn{} \APACrefbtitle {Earth's {{Climate}}: The {{Ocean}}-{{Atmosphere
  Interaction}}} {Earth's {{Climate}}: The {{Ocean}}-{{Atmosphere
  Interaction}}}\ (\BPGS\ 247--259).
\newblock
\APACaddressPublisher{}{{American Geophysical Union (AGU)}}.
\newblock
\begin{APACrefDOI} \doi{10.1029/147GM14} \end{APACrefDOI}
\PrintBackRefs{\CurrentBib}

\bibitem [\protect \citeauthoryear {%
Webster%
\ \protect \BOthers {.}}{%
Webster%
\ \protect \BOthers {.}}{%
{\protect \APACyear {1998}}%
}]{%
webster_monsoons:_1998}
\APACinsertmetastar {%
webster_monsoons:_1998}%
\begin{APACrefauthors}%
Webster, P\BPBI J.%
, Maga{\~n}a, V\BPBI O.%
, Palmer, T\BPBI N.%
, Shukla, J.%
, Tomas, R\BPBI A.%
, Yanai, M.%
\BCBL {}\ \BBA {} Yasunari, T.%
\end{APACrefauthors}%
\unskip\
\newblock
\APACrefYearMonthDay{1998}{{\APACmonth{06}}}{}.
\newblock
{\BBOQ}\APACrefatitle {Monsoons: Processes, Predictability, and the Prospects
  for Prediction} {Monsoons: Processes, predictability, and the prospects for
  prediction}.{\BBCQ}
\newblock
\APACjournalVolNumPages{J. Geophys. Res.}{103}{C7}{14451--14510}.
\newblock
\begin{APACrefDOI} \doi{10.1029/97JC02719} \end{APACrefDOI}
\PrintBackRefs{\CurrentBib}

\bibitem [\protect \citeauthoryear {%
Webster%
, Moore%
, Loschnigg%
\BCBL {}\ \BBA {} Leben%
}{%
Webster%
\ \protect \BOthers {.}}{%
{\protect \APACyear {1999}}%
}]{%
webster_coupled_1999}
\APACinsertmetastar {%
webster_coupled_1999}%
\begin{APACrefauthors}%
Webster, P\BPBI J.%
, Moore, A\BPBI M.%
, Loschnigg, J\BPBI P.%
\BCBL {}\ \BBA {} Leben, R\BPBI R.%
\end{APACrefauthors}%
\unskip\
\newblock
\APACrefYearMonthDay{1999}{{\APACmonth{09}}}{}.
\newblock
{\BBOQ}\APACrefatitle {Coupled Ocean\textendash Atmosphere Dynamics in the
  {{Indian Ocean}} during 1997\textendash 98} {Coupled ocean\textendash
  atmosphere dynamics in the {{Indian Ocean}} during 1997\textendash
  98}.{\BBCQ}
\newblock
\APACjournalVolNumPages{Nature}{401}{6751}{356--360}.
\newblock
\begin{APACrefDOI} \doi{10.1038/43848} \end{APACrefDOI}
\PrintBackRefs{\CurrentBib}

\bibitem [\protect \citeauthoryear {%
Wilks%
}{%
Wilks%
}{%
{\protect \APACyear {2016}}%
}]{%
wilks_stippling_2016}
\APACinsertmetastar {%
wilks_stippling_2016}%
\begin{APACrefauthors}%
Wilks, D\BPBI S.%
\end{APACrefauthors}%
\unskip\
\newblock
\APACrefYearMonthDay{2016}{{\APACmonth{03}}}{}.
\newblock
{\BBOQ}\APACrefatitle {``{{The Stippling Shows Statistically Significant Grid
  Points}}'': How {{Research Results}} Are {{Routinely Overstated}} and
  {{Overinterpreted}}, and {{What}} to {{Do}} about {{It}}} {``{{The Stippling
  Shows Statistically Significant Grid Points}}'': How {{Research Results}} are
  {{Routinely Overstated}} and {{Overinterpreted}}, and {{What}} to {{Do}}
  about {{It}}}.{\BBCQ}
\newblock
\APACjournalVolNumPages{Bull. Amer. Meteor. Soc.}{97}{12}{2263--2273}.
\newblock
\begin{APACrefDOI} \doi{10.1175/BAMS-D-15-00267.1} \end{APACrefDOI}
\PrintBackRefs{\CurrentBib}

\end{thebibliography}

\end{document}